\documentclass[aip,reprint]{revtex4}

\usepackage{graphicx}
\usepackage{verbatim}

\draft % marks overfull lines with a black rule on the right

\usepackage{amsmath}
\usepackage{amsfonts}

\usepackage{color}

\DeclareMathOperator*{\E}{\mathbb{E}}

\begin{document}

\preprint{}

\title{Input nonlinearities can shape beyond-pairwise correlations and improve information transmission by neural populations }

\author{Joel Zylberberg}
%\thanks{}
%\email[]{joelzy@uw.edu}
\affiliation{Department of Applied Mathematics, University of Washington, Seattle WA }

\author{Eric Shea-Brown}
%\thanks{}
\email[]{etsb@washington.edu}
\affiliation{Department of Applied Mathematics, Program in Neuroscience, Department of Physiology and Biophysics, University of Washington, Seattle WA }

\begin{abstract}
While recent recordings from neural populations show beyond-pairwise, or {\it higher-order} correlations (HOC), we have little understanding of how HOC arise from network interactions and of how they impact encoded information.  Here, we show that input nonlinearities imply HOC in spin-glass-type statistical models. We then discuss one such model with parameterized pairwise- and higher-order interactions, revealing conditions under which beyond-pairwise interactions increase the mutual information between a given stimulus type and the population responses. For jointly Gaussian stimuli, coding performance is improved by shaping output HOC only when neural firing rates are constrained to be low. For stimuli with skewed probability distributions (like natural image luminances), performance improves for all firing rates. Our work suggests surprising connections between nonlinear integration of neural inputs, stimulus statistics, %population activity statistics, 
and normative theories of population coding. Moreover, it suggests that the inclusion of beyond-pairwise interactions could improve the performance of Boltzmann machines for machine learning and signal processing applications.%Normative theories might therefore predict differences in the dendritic summation properties in
%populations of neurons with different mean firing rates, or ones adapted (or evolved) to encode stimuli
%drawn from different distributions.

\end{abstract}


\date{\today}

\maketitle

\section{Introduction}

The number of neurons for which activities can be simultaneously recorded is rapidly increasing~\cite{stevenson}. We thus have an advancing understanding of the statistics of population activities, like the relative frequencies of co-active neural pairs, triplets, etc. In particular, much work has investigated the distributions of simultaneously recorded retinal ganglion cell ``words" (patterns of binary neural activities). For some population sizes and stimuli, these distributions are well-fit by pairwise maximum entropy (ME) models~\cite{schneidman06,shlens0609,granot13}, while in other cases beyond-pairwise interactions are evident in the data and models incorporating higher-order correlations (HOC) are needed~\cite{ganmor,tkacik14}. Cortical studies yield similar observations~\cite{yu,ohiorhenuan,montani}.

How do these correlations affect \emph{population coding}? Much work has investigated how pairwise correlations impact a population's ability to transmit information~\cite{averbeck,cafaro,roudi_latham,zohary,tkacik,beck,ecker11,sha04,josic09,hu14}. In particular,~\cite{tkacik} identify optimal pairwise interactions that maximize encoded information in setting very similar to that we will study below.  Coding studies of higher-order correlations (HOC) are limited, but empirical work shows that in some cases, including HOC allows a decoder to recover the stimulus presented to a neural population $3$ times faster than a decoder with access only to pairwise statistics~\cite{ganmor}. Intriguingly, other work~\cite{montani} shows that HOC reduce the mutual information (MI) between the stimuli and resultant population responses.  A recent theoretical investigation~\cite{cayco15} outlines principles by which higher-order correlations can impact the discriminability of pairs of nearby stimuli.  There, second-order response statistics (pairwise correlations) were held fixed at identical values for each stimulus, and the authors showed how triplet correlations could make population responses more discriminable by skewing the distributions of population responses to the two stimuli away from each other.  

%adjusting the third-order statistics can have a substantial effect on the mutual information between stimuli and responses. One possible concern with that study is that a specific range of first- and second-order statistics was explored, leaving it unclear whether the \emph{optimal} population code requires higher-order correlations. For example, could there be some set of first- and second-order statistics not explored by ~\cite{cayco15} that leads to better population codes in the absence of HOC than~\cite{cayco15} were able to obtain with HOC? 

This prior work is intriguing, but begs several questions:   How do HOC impact population coding of multiple stimuli, or continuous families of stimuli -- and not just the stimulus pairs of~\cite{cayco15}?  What network mechanisms can generate the underlying HOC?  
Here, we address both of these questions, identifying when HOC may improve coding performance and how those performance gains come about.  Aside from its relevance for neurobiology, our study suggests that incorporating higher-order interaction terms may improve the performance of Boltzmann machines (including deep belief networks), which are promising algorithms for machine learning applications~\cite{salakhutdinov09,hinton83}

\section{Results}

\subsection{Encoding model} 

We generalize the approach of Tka\v{c}ik \emph{et al.}~\cite{tkacik}, and model the activity of a population of neurons by a triplet-wise ME distribution. Within this model, the stimuli affect neural responses via \emph{bias terms} $h_i =  h^s_i + h^0_i$ (Fig. 1A). Here, $h_i$ is the bias to neuron $i$, $h^0_i$ is the stimulus-independent bias, and $h^s_i$ is the stimulus-dependent bias. As discussed below, the stimuli can be interpreted as additive inputs in a linear-nonlinear neural model, and we define the stimulus distribution by the joint distribution over $h^s_i$~\cite{tkacik}. 

Once the biases are specified, the neural activities~$\{ \vec{\sigma} \}$ ($\sigma_i \in \{0,1\}$ is the silent vs. spiking state of neuron $i$) are distributed as

\begin{eqnarray}
p(\vec{\sigma}| \vec{h})= \frac{1}{Z} \exp \left[\beta \left( \vec{h}\cdot \vec{\sigma} + \sum_{i<j} J_{ij} \sigma_i \sigma_j + \sum_{i<j<k} \gamma_{ijk} \sigma_i \sigma_j   \sigma_k   \right) \right].  
\end{eqnarray} 
\noindent Here, $\beta$ specifies the distribution's width, defining neural reliability~\cite{tkacik,betacomment} analogous to the inverse temperature of an Ising spin-glass model. The parameters $J_{ij}$ and $\gamma_{ijk}$ describe the pairwise interactions and triplet interactions, respectively. The \emph{partition function} $Z = \sum_{\{ \vec{\sigma}\}}   \exp \left[\beta \left( \vec{h}\cdot \vec{\sigma} + \sum_{i<j} J_{ij} \sigma_i \sigma_j + \sum_{i<j<k} \gamma_{ijk} \sigma_i \sigma_j   \sigma_k   \right) \right]$ is the normalizing constant that ensures that all probabilities sum to $1$.  This distribution is the one that specifies the means, covariances, and 3-pt correlations of the activity distribution, while making the fewest possible assumptions about the distribution overall~\cite{jaynes57,amari01,schneidman_connected,schneidman06,shlens0609}. Later, we will optimize over the interaction parameters (and thus the moments of the response distribution), for different stimulus distributions. In so doing, we will identify conditions under which triplet interactions improve stimulus encoding.

%%%%%%%%%Figure%%%%%%%%%%%%%%%%%%
\begin{figure}[tb!]
\begin{center}
\includegraphics[width=4.0in]{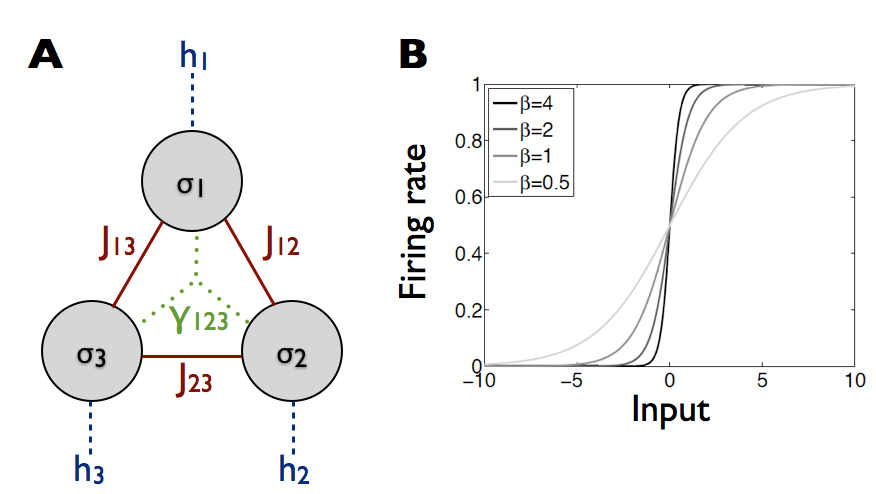}
\caption{(Color online) \textbf{Encoding model.} (\textbf{A}) Each model neuron has a bias $h_i$ (blue) determined by external stimuli.  Recurrent pairwise ($J_{ij}$, red) and triplet ($\gamma_{ijk}$, green) interactions further modify the output statistics.  Illustrative schematic shown for N=3 neurons; models can have arbitrary N. (\textbf{B}) The firing rate (spiking probability) of each neuron varies sigmoidally with the strength of its input $(x_i = h_i + \sum_{j \ne i} J_{ij} \sigma_j +  \sum_{j<k; j,k \ne i} \gamma_{ijk}  \sigma_j \sigma_k )$; see text. The steepness of the sigmoid depends on $\beta$.}
\end{center}
\end{figure}
%%%%%%%%%Figure%%%%%%%%%%%%%%%%%%

As emphasized by \cite{tkacik}, this parameterization of  $p(\vec{\sigma}| \vec{h})$ can be interpreted as a static nonlinear input-output neural model, within a network with symmetrical connections between units. Consider the probability of one neuron firing (having $\sigma_i=1$), conditioned on the states of the other neurons and the biases:
\begin{eqnarray}
p(\sigma_i = 1| \{\sigma_{j\ne i} \}, \vec{h} ) =  
	g[\beta( h_i + \sum_{j \ne i} J_{ij} \sigma_j + \sum_{j<k; j,k \ne i} \gamma_{ijk}  \sigma_j \sigma_k )],
\end{eqnarray}
where $g(\beta x_i) = (1 + e^{-\beta x_i})^{-1}$ is a sigmoidal function (Fig. 1B). To obtain Eq. 2 from Eq. 1, consider the conditional probability distribution $p(\sigma_i | \{\sigma_{j\ne i} \}, \vec{h} ) $, where $\sigma_i$ can take on one of two values: either $0$ or $1$. The exponential ``Boltzmann factors" for each of the two states, obtained by using $\sigma_i = 0$ or $1$ in Eq. 1, while keeping everything else fixed, are $\exp \left[\beta \left( \sum_{j \ne i} \sigma_j h_j + \sum_{j<k; j,k \ne i} J_{jk} \sigma_j \sigma_k + \sum_{j<k<l; j,k,l \ne i} \gamma_{jkl} \sigma_j \sigma_k   \sigma_l   \right) \right]$ and $\exp \left[\beta \left(h_i + \sum_{j \ne i} J_{ij} \sigma_j + \sum_{j<k j; j,k \ne i} \gamma_{ijk}  \sigma_j   \sigma_k   \right) + \beta \left( \sum_{j \ne i} \sigma_j h_j + \sum_{j <k; j,k\ne i} J_{jk} \sigma_j \sigma_k + \sum_{j<k<l; j,k,l \ne i} \gamma_{jkl} \sigma_j \sigma_k   \sigma_l   \right) \right]$ respectively. Summing these two Boltzmann factors, we get the conditional partition function, and dividing the appropriate Boltzmann factor by this conditional partition function, we obtain Eq. 2.

The firing probability in one discrete time bin is akin to the mean firing rate. Since firing rates that vary sigmoidally with synaptic input are commonly encountered~\cite{maass99,brillinger,ho2000}, we interpret the argument ($x_i$) of the sigmoid as the input to a linear-nonlinear model neuron. 
With no beyond-pairwise interactions ($\gamma_{ijk} = 0$), the bias $h_i$ and recurrent inputs to the neuron $\{J_{ij} \sigma_j \}$ add; the sigmoidal function of that sum determines the firing rate. If $\gamma_{ijk}>0$, then when neurons $j$ and $k$ are co-active, the recurrent input to neuron $i$ is $J_{ij} + J_{ik} + \gamma_{ijk}$, which is larger than the sum of the contributions observed when only one recurrent input is active at a time ($J_{ij}$+ $J_{ik}$); these inputs combine super-linearly. Conversely, for $\gamma_{ijk}<0$, they combine sub-linearly. Thus, the way that synaptic inputs combine maps onto triplet interactions in statistical models of population activity, shaping beyond-pairwise correlations. If the recurrent input to neuron $i$ is an arbitrary function of the activities of the other neurons, $x_i = h_i + f(\{\sigma_{j\ne i}\})$, triplet interactions come from the first nonlinear terms in the series expansion of $f(\cdot)$ (see Appendix A). 

We note that, in this mechanistic interpretation of our probability model, both  the recurrent connections $J_{ij}$, and the super- or sub-linear integration (given by $\gamma_{ijk}$) are symmetrical in their indices. While this symmetry is physically realizable in neuronal networks, it is not the most general possible configuration. We will later return to possible biophysical mechanisms behind such nonlinearities. We further note that, in our model, these ``recurrent" interactions are instantaneous, which is not true for physical neurons. Studies of the present mechanism for HOC in recurrent dynamical models, as in Ref.~\cite{andrea}, are an intriguing area for future work.  Finally, we note that, in addition to recurrent coupling, interaction terms can also come from common noise inputs to multiple neurons~\cite{macke,andrea,leen}, and that in this case, the interactions are likely to be symmetrical, as in Eq. 1.

\subsection{When do HOC improve coding?: -- analytical results} 

Having motivated our probability model, we ask  when triplet interactions improve coding. To do this, we use the framework introduced by Tka\v{c}ik \emph{et al.}~\cite{tkacik} to study population coding with pairwise interactions. For a given stimulus distribution and parameter set $\{\beta, \{h^0_i\}, \{J_{ij}\}, \{\gamma_{ijk} \} \}$, we compute the mutual information (MI) between stimuli $\vec{h^s}$ and the responses $\vec{\sigma}$:
\begin{eqnarray}
MI = - \sum_{ \{ \vec{\sigma} \}} p(\vec{\sigma}) \log[  p(\vec{\sigma})  ]   + \int d \vec{h^s} p(\vec{h^s}) \sum_{ \{ \vec{\sigma} \}} p(\vec{\sigma}| \vec{h^s}) \log[  p(\vec{\sigma}| \vec{h^s})  ].  
\end{eqnarray}
The first term is the \emph{response entropy}, and the second term is (minus) the mean entropy of the response conditioned on the stimulus (\emph{noise entropy}). 

We will first discuss analytical results, obtained in the limit of weak stimulus-dependence of the neural responses -- corresponding to small $\beta$ -- for arbitrary $\{ \{h^0_i\}, \{J_{ij}\}, \{\gamma_{ijk} \} \}$. We will then show numerical results indicating that our qualitative findings persist over a range of stimulus-coupling strengths. The analytical calculations are quite tedious, and so we describe them briefly here, and show all the details in Appendix B. For the analytical investigation, we re-write our probability distribution as

\begin{eqnarray}
p(\vec{\sigma}| \vec{h})= \frac{1}{Z} \exp \left[\epsilon \vec{h^s}\cdot \vec{\sigma} + \beta \left( \vec{h^0}\cdot \vec{\sigma} +  \sum_{i<j} J_{ij} \sigma_i \sigma_j + \sum_{i<j<k} \gamma_{ijk} \sigma_i \sigma_j   \sigma_k   \right) \right],  
\end{eqnarray} 
where $\epsilon$ parametrizes the strength of the stimulus-dependence of the neuronal responses~\cite{expansion_comment}. We then expand the MI (Eq. 3) in powers of $\epsilon$. This expansion yields

\begin{eqnarray}
MI &=&   \frac{\epsilon^2}{2} \left[     \sum_i  \left< \left(h^s_i\right)^2 \right> \text{var}(\sigma_i) + \sum_{i\ne j} \left<h^s_i h^s_j \right> \text{cov}(\sigma_i,\sigma_i) \right]   \\
&+& \epsilon^3 \left[ \sum_{i} \frac{\mu_i}{3} \left(1 + 2 \mu^2_i - 3\mu_i \right) \left< h^3_i \right> +  \sum_{i\ne j} \text{cov}(\sigma_i,\sigma_j)\left(1 - 2 \mu_i \right) \left< h^2_i h_j  \right>  \right] \nonumber \\
&+& \epsilon^3 \left[ \sum_{i \ne j \ne k} \frac{1}{3} \left( \tau_{ijk} + 2 \mu_i \mu_j \mu_k - 3 \mu_k \pi_{ij} \right) \left<h_i h_j h_k \right>   \right]  \nonumber \\
&+& \mathcal{O}(\epsilon^4), \nonumber
\end{eqnarray}
where $\mu_{i} = \E \left[ \sigma_i | \vec{h^s} = \vec{0} \right]$, $\pi_{ij} =  \E \left[ \sigma_i \sigma_j| \vec{h^s} = \vec{0} \right]$, $\tau_{ijk} =  \E \left[ \sigma_i \sigma_j \sigma_k| \vec{h^s} = \vec{0} \right]$, angled brackets denote expectations over the stimulus distribution, and $\text{cov}(\sigma_i,\sigma_i) = \pi_{ij} - \mu_i \mu_j$ and $\text{var}(\sigma_i) = \pi_{ii} - \mu_i^2$ are moments of the \emph{spontaneous activity distribution} (obtained when no stimulus is present, and thus $h^s_i = 0~\forall~i$). 

The form of Eq. 5 emphasizes that the MI depends on the relationships between the moments of the stimulus distribution, and the moments of the response distribution in a very specific way. In particular, it is the moments of the \emph{spontaneous activity distribution} (obtained when no stimulus is present, and thus $h^s_i = 0~\forall~i$) that determine the MI. This effect arises because the Boltzmann factors (the exponentials in the numerator of) Eq. 4 factorize into a stimulus-dependent part $\exp( \epsilon \vec{h^s} \cdot \vec{\sigma})$, and a stimulus-independent part. In the case where the stimulus is set to zero, the stimulus-dependent part is unity, and one obtains the Boltzmann factors for the spontaneous activity distribution. Thus, the Boltzmann factors consist of a stimulus-dependent term multiplied by the Boltzmann factors for the spontaneous activity distribution. Consequently, the moments of the spontaneous activity distribution determine the statistical interactions of the encoding model, and thus the MI. Our analytical expansion exploits this multiplicative structure to obtain Eq. 5.

Because Eq. 5 yields the MI as a function of moments of the spontaneous activity distribution -- and the interaction terms in Eqs. 1 and 4 determine those moments -- we can use Eq. 5 to understand the conditions under which different interactions will improve MI.  

Our first observation is that, in the case of stimulus distributions that are symmetric about their mean (like, for example, Gaussian distributions), the odd moments vanish, and thus so do the $\mathcal{O}(\epsilon^3)$ terms in Eq. 5. In this case, at least up to $\mathcal{O}(\epsilon^4)$, the MI depends only on the first and second moments of the response distribution. Consequently, in this case, adjusting the HOC (by having $\gamma \ne 0$) will not impact the MI (at least up to $\mathcal{O}(\epsilon^4)$). Because the inclusion of HOC lowers the response entropy -- which can reduce the MI -- we conjecture that the optimal population code in this case will have no HOC. This conjecture is supported by our numerical investigation (below, Figs. 3AC).

A second prediction is that,  for this case of symmetric stimulus distributions, the best MI is obtained when the pairwise interactions between neurons -- which determine the signs of $\text{cov}(\sigma_i,\sigma_i) $ in Eq. 5 -- have the same signs as the correlations between their stimulus-dependent gains $\left<h^s_i h^s_j \right>$. This finding is consistent with the numerical results obtained by ~\cite{tkacik}, in the case of small $\beta$ (which, in their work, parametrizes both the strength of the stimulus coupling and of the interactions between cells). We note that, because our analytical expansion (Eq. 5) holds only in the limit of small $\beta$, we cannot use this expansion to understand the numerical results that~\cite{tkacik} obtained in the limit of large~$\beta$.

 In the case of skewed stimulus distributions, the $\mathcal{O}(\epsilon^3)$ terms in Eq. 5 are non-zero, and thus changes to the HOC -- by adjusting the third-order interaction $\gamma$ --  can increase the MI: this is confirmed by the numerical results shown in Fig. 3 BDF (below). Unfortunately it is not straightforward to determine the optimal sign of $\gamma$ from Eq. 5 in the case of unskewed stimulus distributions, because triplet moments in the response distribution will be simultaneously determined by the $J$ and $\gamma$. However, Eq. 5 does allow us to understand how HOC might impact coding in large neural populations: in a population of $N$ neurons, there are $\mathcal{O}(N^3)$ of the cubic terms in Eq. 5 (that depend on HOC), but only $\mathcal{O}(N^2)$ of the quadratic terms (that do not depend on the HOC). Consequently, in the limit of large $N$, the HOC could have a large impact on the MI.

\subsection{When do HOC improve coding?: -- numerical results} 

Our analytical investigation identified situations -- of skewed stimulus distributions -- in which we expect HOC to enhance the population code. To verify this finding (and investigate the extent to which it applies to stronger stimulus-coupling regimes than could be considered by our series expansion),  we numerically seek the $h^0_i,~J_{ij}$, and $\gamma_{ijk}$ in Eq. 1 that maximize the MI (Eq. 3). We repeat this optimization for different values of $\beta$ -- and thus different strengths of stimulus coupling: note that, in Eq. 1, $\beta$ multiplies the stimulus, and thus it sets the strength of the stimulus coupling.

To simplify our numerical investigation, we consider homogeneous parameter values: $h^0_i = h^0$,  $J_{ij} = J$, and  $\gamma_{ijk} = \gamma~\forall i,j,k$. For consistency with this, we use permutation-symmetric stimulus distributions. However, for any given stimulus example $\vec{h^s}$, the conditional response distribution will not necessarily be permutation-symmetric. %We leave the non-permutation symmetric case for future work. 
For a given set of model parameters, we numerically compute the MI using Monte Carlo methods, and we optimize over the parameters using gradient ascent (see Appendix C).

As a control, to see if and when the HOC really do lead to better MI than could be obtained by a model with no higher-order correlations, we also optimize MI over $h^0$ and $J$ while imposing the constraint $\gamma=0$, the {\it triplet forbidden} case. In this case, the conditional response distribution becomes the pairwise maximum entropy model, which is the situation considered by~\cite{tkacik} (here, with homogeneous parameters). Comparing the maximum attainable MI with triplet interactions allowed or forbidden, we ascertain when, and how much, their presence improves coding. This is related to $3^{rd}$ order {connected information}~\cite{schneidman_connected}, where one fits both $2^{nd}$ and $3^{rd}$ order ME models to the stimulus-conditioned response distributions and compares the resulting MI. In the case of~\cite{schneidman_connected} (and in the recent work of~\cite{cayco15}), the second-order moments are held fixed as the third-order moments are varied, whereas we separately \emph{optimize} the interaction parameters controlling those moments in the two cases. Because we separately optimize the $2^{nd}$ and $3^{rd}$ order models, we obtain more conservative estimates of how MI increases due to the $3^{rd}$ order interactions than did~\cite{cayco15}.

We note that, when triplet interactions are allowed, the optimum can still occur for $\gamma=0$.  In this case the maximal MI will be equal for networks allowing and forbidding triplet interactions. Thus, because the triplet-allowed model space is a superset of the triplet-forbidden one, it will never be the case that the optimized triplet-allowed network performs worse than the optimized triplet-forbidden one.

We begin with the case of jointly Gaussian stimuli (Fig. 2A) of varying levels of correlation $\rho$. Recall that our analytical results (Eq. 5, above) suggest that, in this case, it is only the second-order moments of the response distribution that affect the MI, and that we conjectured based on this that HOC confer no coding benefit. This is borne out by our numerical analysis: (Fig. 3A,C, $N=10$ neurons):  even when triplet interactions are allowed, the optimal encoder has $\gamma=0$ (Fig. 4A). Here, the stimulus distribution is symmetric about the mean, and the optimal encoder is on-off symmetric with $\left< \sigma \right> = 0.5$. That symmetry -- with neurons being in either the ``spiking" (1) or ``non-spiking" state (0) with equal probability -- maximizes response entropy. When the stimuli are drawn from discrete binary distributions with equal probabilities for the two states, we also find that triplet interactions confer no coding advantage (data not shown). These observations support our analytical finding that, for unskewed stimulus distributions, with on-off symmetric response distributions, triplet interactions are not useful for coding. Moreover, these numerical results hold over a range of $\beta$ values, and thus a large range of stimulus-coupling strengths (and not just the weak ones considered in our analytical calculation).

To seek situations when triplet interactions might be beneficial to the population code, we break the on-off symmetry. We do this in two different ways, the first of which is to consider skewed stimulus distributions. The case of skewed stimulus distributions is potentially important because the distribution of membrane potentials (neuronal inputs) in rat auditory cortical neurons is skewed~\cite{deweese_zador}, as are the spike-count distributions in dichotomized Gaussian models of neural population activity~\cite{macke}. Moreover, naturalistic stimuli (like pixel values in real-world images) have skewed distributions (Fig. 2BC). Consequently, the inputs to real biological neurons -- either from external stimuli, or from other neurons within the brain -- may show nonzero skew~\cite{skew_comment}. We use as stimuli calibrated luminance images from the database of Tka\v{c}ik \emph{et al.}~\cite{tkacik_images}. 

It is worth noting that images of natural scenes (as one might collect with a digital camera in, say, a forest) have statistical properties that differ markedly from purely random white-noise images. Much theoretical and empirical work has investigated these properties~\cite{simoncelli_olshausen,barlow,jz_pfau,stephens_image_statmech,ruderman97,ruderman_scaling,hsiao}. Of particular note are the rich correlation structure, and the power-law power spectra: natural images have autocorrelation functions (Fourier transform of the power spectra) that decrease with distance~\cite{ruderman_scaling}, in a manner that is surprisingly independent of the occlusion property of objects in those images~\cite{hsiao, jz_pfau}.  
By drawing groups of pixels (Fig.~2) with variable spacing $d$, we vary the level of correlation between stimulus values. Since luminance (or photon count) is non-negative, but can be arbitrarily large, this distribution is skewed (Fig.~2C).  %This skew is not unique to luminance stimuli. For example, the distribution of membrane potentials in rat auditory cortical neurons is skewed~\cite{deweese_zador}, as are the spike-count distributions in dichotomized Gaussian models of population activity~\cite{macke}. Consequently, the inputs to real biological neurons -- either from external stimuli, or from other neurons within the brain -- may show nonzero skew.

%%%%%%%%%Figure%%%%%%%%%%%%%%%%%%
\begin{figure}[t!]
\begin{center}
\includegraphics[width=5in]{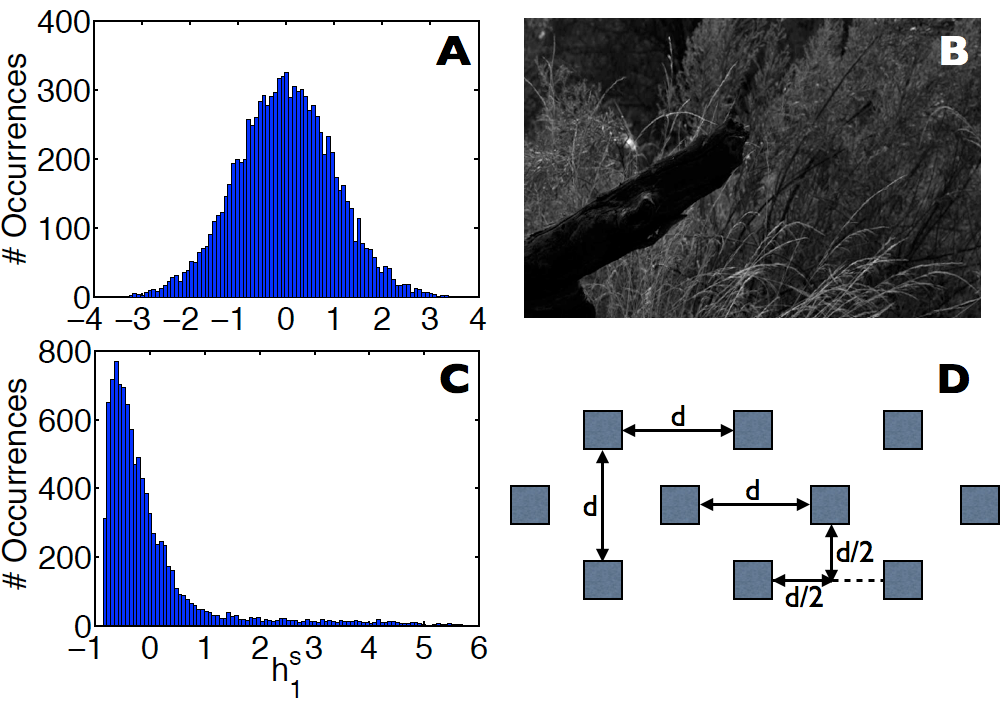}
\caption{(Color online) \textbf{Stimuli considered in this work.} (\textbf{A}) Marginal histogram of one stimulus, $h^s_1$, for the jointly Gaussian ensemble. (\textbf{B}) One of the natural luminance images used in this work, from the database of Tka\v{c}ik et al.~\cite{tkacik_images}. The marginal histogram of pixel values for this set of images (normalized to have zero mean and unit variance) is skewed (\textbf{C}). To generate our naturalistic stimulus ensemble, we randomly draw dectuplets of pixel values with spacing $d$ pixels by placing the template (\textbf{D}) at a random location on a randomly chosen image, and setting stimulus values $h^s_i$ to match the pixel values falling under each square. To maintain permutation symmetry of our stimulus ensemble, we permute which square corresponds to which stimulus index $i$ for each draw. The marginals (\textbf{A,C}) are the same for all stimulus indices $i$.}
\end{center}
\end{figure}
%%%%%%%%%Figure%%%%%%%%%%%%%%%%%%
For the natural image luminance stimuli, we find that triplet interactions indeed confer a coding advantage.  For $N=10$ cells, this is a $5$ --- $10 \%$ improvement in MI compared to the optimized purely pairwise encoder (Figs. 3B,D).  The advantage is largest for close-by sampled pixels (small $d$), and at relatively low values of $\beta$ (i.e., relatively unreliable neurons). Natural images have rich beyond-pairwise statistics~\cite{simoncelli_olshausen}. Is that why triplet interactions improve encoding for natural image stimuli? No: repeating these optimization experiments using linear mixtures of variables from skewed Pearson-system marginal distributions as stimuli, we also observed that triplet interactions improved coding (data not shown). These results support our analytical findings that, in the case of skewed stimulus distributions, triplet interactions can improve coding performance.

While our analytical results focused on the statistics of the stimulus distribution, one can also break the on-off symmetry of our problem by restricting the firing rates (FR's) of the neurons in our networks.  Thus far, they have been allowed to take arbitrary values.  Empirically, however, neurons are seen to fire infrequently~\cite{hromadka,baddeley,schneidman06,ohiorhenuan,ganmor}, with mean FR's of a few Hz: for 10 --- 20 ms time bins~\cite{ohiorhenuan,ganmor,schneidman06} this yields $\left< \sigma \right> \sim 0.01$ --- $0.1$. To incorporate this constraint, we maximized a Lagrange function $\mathcal{L} = MI - \lambda \left< \sigma \right>$ that disfavors high FR's~\cite{tkacik,karklin_simoncelli}, similar to the notion of \emph{sparse coding}~\cite{karklin_simoncelli,jz_plos,olshausen96,kzd,jz_sailsparse}. By varying $\lambda$, we alter the mean FR of the optimal network~\cite{tkacik}. We ask how the MI for these optimal networks varies as a function of their mean FR for networks with triplet interactions either allowed or forbidden. For these investigations, we restricted ourselves to $\beta = 1.5$. 

Intriguingly, for jointly Gaussian stimuli triplet interactions improve coding performance at sufficiently low firing rates (Fig. 3E). The improvement is larger for stronger stimulus correlations. This effect is somewhat surprising because our analytical results (Eq. 5) show that, up to $\mathcal{O}(\epsilon^3)$, MI is not increased by the inclusion of HOC when the stimulus distribution is unskewed. We thus ascertain that the results in Fig. 3E arise at higher-order terms in the series expansion. The key intuition is that, when the encoding problem is sufficiently symmetric (i.e.: the stimulus distribution is symmetric about the mean \emph{and} no firing rate constraint prevents the response distribution from being on-off symmetric), the best encoders have no triplet interactions. 

For natural image pixel stimuli (with skewed distributions), restricting firing rates lead to further benefits of triplet interactions, over and above those already seen with unconstrained firing rates.  In particular, for low firing rates the benefits of triplet interactions can be as large as $15$ --- $ 20 \%$ (Fig. 3F).  %We note again that that larger effect sizes may be possible in other circumstances -- like those with larger population sizes -- not considered herein (for example, see discussion of analytic calculations above).

The results shown herein are for networks of $N=10$ neurons. This is as large as we can consider while being able to numerically optimize our MI function with reasonable speed (see Appendix C for methods). However, we emphasize that our analytical results (Eq. 5) hold for arbitrary $N$, and in fact suggest the effect of higher-order interactions on MI may grow with population size (see discussion of analytic calculations above). 

%%%%%%%%%Figure%%%%%%%%%%%%%%%%%%
\begin{figure}[t!]
\begin{center}
\includegraphics[width=5in]{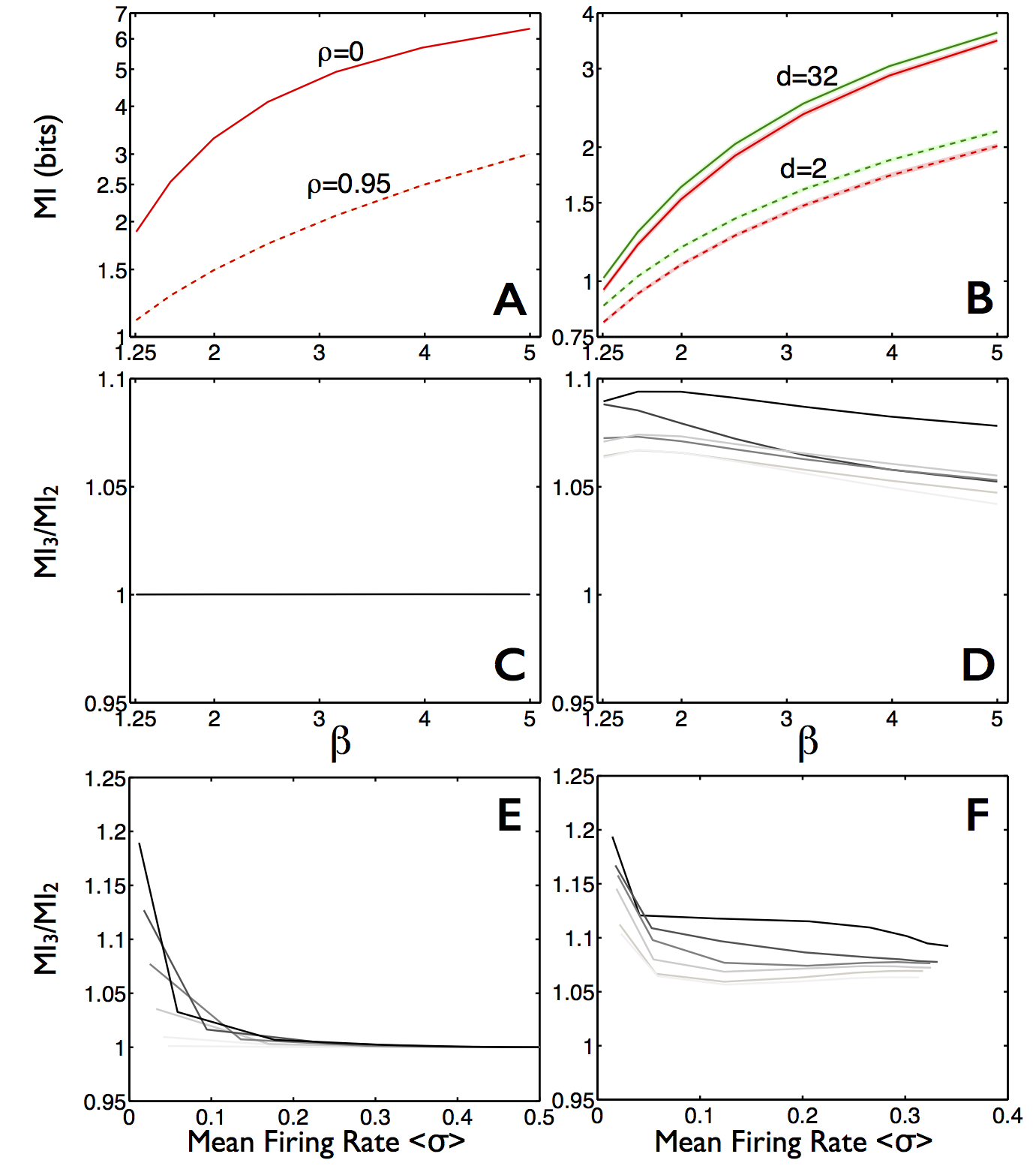}
\caption{(Color online) \textbf{Triplet interactions can improve encoding of stimuli.} Panels (\textbf{ACE}) show results for Gaussian-distributed stimuli, whereas panels (\textbf{BDF}) are for natural image pixel stimuli, which have skewed distributions. (\textbf{A}) For jointly Gaussian stimuli with pairwise correlation coefficients of 0.95 (dashed lines) or 0 (solid lines), encoders with triplet interactions allowed (green) or forbidden (red, $\gamma=0$) have the same coding performance, which increases with neural reliability $\beta$. (\textbf{B}) For natural image stimuli with pixel spacings of $d=32$ (solid lines) or $d=2$ (dashed lines), the triplet-allowed encoder (green) performs better. The shaded regions (similar in thickness to the lines) around the lines in panels (\textbf{A,B}) show the standard deviation of the mean MI over 5 repeats of the optimization procedure with different sets of random stimuli. (\textbf{C,D}) To summarize how performance gains vary with correlation level, we plot the ratio of the MI for the optimal triplet-allowed networks ($MI_3$) to the one for triplet-forbidden networks ($MI_2$, $\gamma=0$) as a function of $\beta$. The darkest curve corresponds to the largest correlation ($\rho=0.95$ for the Gaussian (\textbf{C}) and $d=2$ for the natural image stimuli (\textbf{D})), the lightest curve corresponds to the smallest correlation ($\rho=0$ for the Gaussian and $d=32$ for the natural image stimuli), and intermediate shades correspond to intermediate levels of correlation. (\textbf{E,F}) For $\beta=1.5$, we similarly plot the performance ratio as a function of mean firing rate for Gaussian (\textbf{E}) and natural image (\textbf{F}) stimuli, in cases with constrained firing rates (see text). In order to make a fair comparison, we estimate the MI of the triplet-allowed and triplet-forbidden networks at the same firing rate (see Appendix C for details).  $N=10$ neurons for all cases.}
\end{center}
\end{figure}
%%%%%%%%%Figure%%%%%%%%%%%%%%%%%%

%%%%%%%%%Figure%%%%%%%%%%%%%%%%%%
\begin{figure}[t!]
\begin{center}
\includegraphics[width=4.5in]{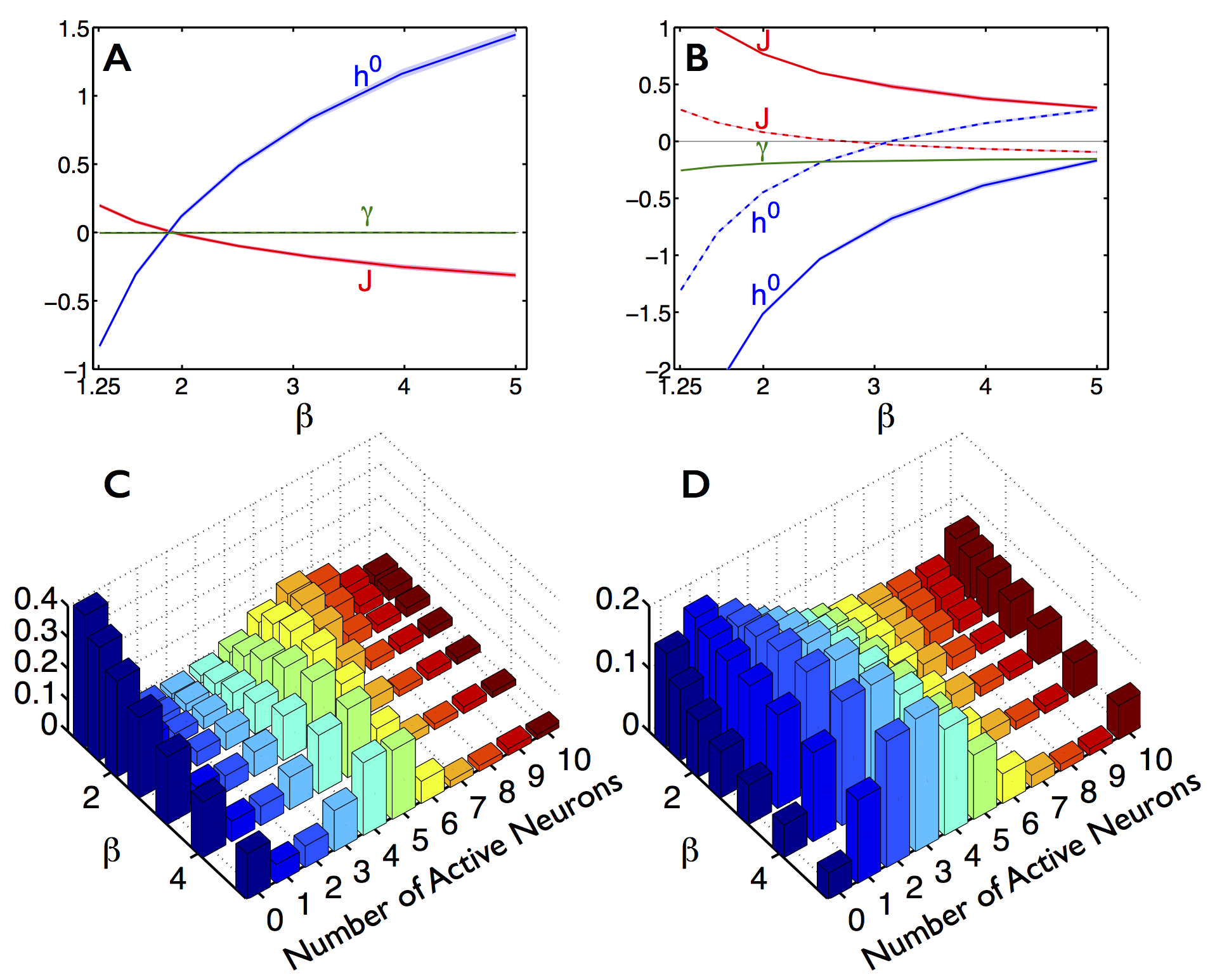}
\caption{(Color online) \textbf{Triplet interactions, when they are beneficial, sparsify the neural representation of stimuli.} 
(\textbf{A}) For jointly Gaussian stimuli ($\rho=0.95$) and no firing rate constraint, optimal encoders have no triplet interaction ($\gamma$, green). At low $\beta$, these optimal encoders have pairwise interactions ($J$, red) that enhance the input correlations, whereas at high $\beta$ they oppose them~\cite{tkacik}. The optimal biases $h^0$ (blue) cancel the effective mean-field bias from pairwise interactions, $h_{\text{eff}} = (N-1)J \left< \sigma \right> $. (\textbf{B}) For the natural image pixel stimuli (d=2), the optimal encoders have negative triplet interactions (when triplet interactions are allowed: solid lines) for all $\beta$  -- the encoder parameters do not change sign. When triplet interactions are forbidden (dashed lines), the magnitudes of the pairwise interaction and bias are smaller, and do change sign with increasing $\beta$. The solid horizontal line (at zero) is to guide the eye. The shaded regions (similar in thickness to the lines) around the lines in panels (\textbf{A,B}) show the standard deviation of the mean parameter values over 5 repeats of the optimization procedure with different sets of random stimuli. For other levels of stimulus correlation, we find qualitatively similar results (not shown). A comparison of the response distributions of the optimal encoders for the natural image pixel stimuli (d=2) with triplet interactions allowed (\textbf{C}) or forbidden (\textbf{D}, $\gamma=0$) shows that the triplet interactions sparsify the responses by reducing the probability of the state in which all neurons are active.}
\end{center}
\end{figure}
%%%%%%%%%Figure%%%%%%%%%%%%%%%%%%

\subsection{How do HOC improve coding?} To understand how the HOC improve the population code, we investigated the parameters of the optimized models: the $J$, $h^0$, and $\gamma$ that maximized the MI in different cases.

For skewed natural image pixel stimuli, the negative ($\gamma<0$) triplet interactions we observed at optimality (Fig. 4B) {\it sparsify} neural responses by reducing the frequency of multi-spike synchrony in which many neurons fire simultaneously.  This sparsifying role of triplet interactions agrees with experimental findings~\cite{ohiorhenuan} and mechanistic modeling~\cite{macke,amari03,andrea}.  Importantly, $\gamma<0$ is optimal even in the absence of a FR constraint, pointing to a richer role in shaping response distributions.  %With no FR constraint it is unclear why such sparsifying is beneficial. 

Following~\cite{tkacik}, we first note that (at least at small $\beta$ when $\gamma=0$, and for all $\beta$ when $\gamma \ne0$) the optimal encoders have positive $J$ (Figs. 4A,B).  Thus, pairwise network interactions reinforce the positive correlations already present in the stimulus, which~\cite{tkacik} interpret as an error-reducing property:  responses tend to be constrained to a smaller set of possibilities.  This effect can be beneficial only up to a point:  for very large positive $J$, neurons would all fire synchronously regardless of the stimulus, sharply reducing response entropy. There is therefore a trade-off between the desiderata of error reduction ($J$ reinforces correlations) and high response entropy ($J$ opposes correlations). %The balance between these determines the optimal $J$ value~\cite{tkacik}. 

Triplet interactions impact the tradeoff in a novel way:  $\gamma<0$ combats multi-spike synchrony, so that response entropy can be maintained even with $J$ reinforcing stimulus correlations~\cite{MEcomment}. Triplet interactions are more suited than pairwise ones at specifically suppressing multi-spike states, in line with the observation that $\gamma<0$ and $J>0$ at optimality, and not vice versa (see Appendix D).   The response distributions of optimal encoders with triplets allowed (Fig. 4C) or forbidden (Fig. 4D) support this notion:  even with no constraints on the FR, the triplet-allowed network makes less use of the state in which all neurons are active.
%, consistent the notion that the tendency towards synchrony limits performance when triplet interactions are forbidden, and that the triplet interactions combat that effect. 
Moreover, with triplet interactions forbidden, the optimal encoders have smaller $J$ (Fig. 4B), also consistent with the interpretation above.  Finally, we observed $\gamma<0$ to be optimal when we constrained the firing rates as well (data not shown).
%; here, this allows the network to maintain positive pairwise interactions (J), and the corresponding error-correcting properties, while still preferentially using states in which relatively small numbers of cells are co-active. 

Allowing nonzero triplet interactions yields optimized network parameters $J$ and $\gamma$ that do not change sign as $\beta$ is varied. This stands in contrast to the case of $\gamma=0$, for which the encoder parameters change sign as $\beta$ is varied (Fig. 4): at low $\beta$, they reinforce the stimulus correlations, while at high $\beta$, they oppose them. This behavior is dictated by the trade-off between noise and response entropies described above~\cite{tkacik}.

\section{Discussion and Conclusions}
We have shown that, in the case of skewed stimulus distributions, or of low neural firing rates, neural population coding efficacy can be enhanced by third-order correlations between neurons.

\subsection{What about $4^{th}$ order and higher interactions?} 

While we have herein restricted ourselves to third order interactions, the same methods would apply equally well to higher-order models (with $4^{th}$ or $5^{th}$, or higher-order terms in the exponential in Eq. 1). This naturally begs the question: ``What order is high enough?'' In other words, what is the highest order of interactions that one must consider in order to understand information transmission in neural systems?  Evidence from available experiments seems to suggest an encouraging answer.  For recordings of $\approx 100$ cells from retina being stimulated with naturalistic movies~\cite{ganmor} it appears that, even at 4th order, the number of non-zero interactions is quite small, and by sixth order, there are none.  Recordings from somatosensory cortex \cite{montani} also suggest that the order of needed interactions will be much smaller than the recorded number of neurons.  To summarize, the type of approach used in this paper could be extended to higher orders, and available experiments suggest that such a venture will have a well-defined stopping point: once we understand how interactions up to order 5 or 6 affect coding, we could be largely done.

\subsection{Potential biophysical origins of beyond-pairwise interactions} 
 
 In our model, higher-order interactions can arise from the nonlinear combination of recurrent inputs. Neurobiology provides several processes which can affect such nonlinear combinations.  
Even for passive single-compartment neurons (no dendrites), inputs can combine sub-linearly, as follows.  Synaptic inputs open ion channels, moving the membrane potential towards that ion's reversal potential~\cite{dayan_abbott}. Opening subsequent channels creates less current as there is less driving force pushing ions through the channel~\cite{gulledge}. Dendrites have additional properties that yield  nonlinearities~\cite{gulledge,mel,harnett,london}. This allows flexible higher-order interactions:  both super- and sub-linear dendritic summation are observed when two inputs impinge on the same dendritic branch~\cite{polsky}, while inputs to separate branches combine linearly.  For strong  dendritic inputs, the observed integration properties are sub-linear~\cite{polsky}, corresponding to negative triplet interactions, similar to what we observed (Fig. 4) for optimal coding.  

We emphasize that, in our probability model, the nonlinear integration is symmetric for the cell triplets (Eq. 1). In other words, if neuron i super-linearly combines the inputs from neurons j and k, then neuron j super-linearly combines the inputs from cells i and k, etc. While this is a highly constrained case, it is physically realizable. At the same time, the nonlinear statistical interactions in our probability model can also arise from common noise inputs to multiple neurons~\cite{macke,andrea,leen}. In that case, the interactions will naturally be somewhat symmetrical.

%This is also in intriguing \Ecomment{I am probably overusing that word my now} agreement with experimental measurements of beyond-pairwise interactions between nearby cells~\cite{ohiorhenuan}.    \Ecomment{again cite Yu here if relevant}

%Herein, we considered  nonlinear combinations of recurrent inputs. Including $3^{rd}$ order terms like $h_i \sigma_i \sigma_j$ in our log-polynomial probability distribution (Eq. 1), one can model other input  nonlinearities.  %One avenue for future work is to study biophysical neuron models, and ask whether  passive and active  nonlinearities improve coding like we have seen in our statistical model.  

We note that caution is warranted when making mechanistic interpretations of statistical parameters observed in neural data.  Pairwise interactions in those data do not necessarily reflect synaptic couplings, as there may be common input to both neurons from unobserved cells that are the cause the correlation. Similar remarks apply to higher-order interactions, which can also be driven by ``hidden" (unobserved) cells~\cite{urs}, spike-generating nonlinearities~\cite{macke,amari03,andrea} and other mechanisms~\cite{schneidman_connected,andrea}.  

\subsection{Implications for machine learning and computer vision}  

In both the mammalian visual system, and in Boltzmann-type computer vision algorithms~\cite{salakhutdinov09,hinton83}, continuous-valued pixel intensities are encoded into binary. These codewords describe the states of all computational units in the system, each of which are either ``on", or ``off" at any given moment. This binarization naturally leads to a loss of information about the image, potentially hindering these systems' abilities to perform the hard computational task of object recognition. Herein, we investigated the features that can minimize the information loss in binary image processing systems. Our results indicate that pairwise and beyond-pairwise statistical interactions between computational units can improve the performance of Boltzmann-type image encoders, like deep belief networks~\cite{salakhutdinov09}.  In other words, we identify and begin to explain an information-theoretic role for recurrent interactions {\it within} such network layers.  At the same time, these higher-order interactions might be difficult to optimize (i.e., to find suitable learning rules).

\subsection{Summary and Conclusions} 

 We have demonstrated that input nonlinearities can generate beyond-pairwise interactions in spin-glass statistical models of neural population activity, and observed that -- under biologically relevant conditions -- these nonlinearities can improve population coding. In particular, we find that the third-order interactions can improve coding when the stimulus distribution is skewed, and/or when the neurons are restricted to have reasonably low firing rates. Normative theories might thus predict differences in the summation properties of neurons in networks that are evolved (or adapted) to encode different types of stimuli, or in networks with different pressures to regulate firing rates.

%%%%%%%%%%%%%%%%%%%%%%%%%%%%%%%%%%%%%%%%%%%%%%%%%%%
%%%%%%%%%%%%%%%%%%%%%%%%%%%%%%%%%%%%%%%%%%%%%%%%%%%

\acknowledgments
\smallskip
\noindent We thank  N. Alex Cayco Gajic for motivation and helpful insights regarding this work, and N. Alex Cayco Gajic as well as Braden Brinkman, Jakob Macke, and Mike DeWeese for comments on the manuscript. This work was supported by NSF Career grant DMS-1056125 and a Simons Fellowship in Mathematics (to ESB). 
\appendix

\section{Triplet interactions arise from the first non-linear terms in a series expansion of the input to our model neuron} 
\label{sec:triplet_expansion}

If we let the recurrent input to neuron $i$  be an arbitrary function of the activities of the other neurons, $x_i = h_i + f(\{\sigma_{j\ne i}\})$, triplet interactions arise from the first nonlinear terms in the series expansion $x_i = h _i + \sum_{j\ne i} a_{ij} \sigma_j + \sum_{j \ne i} b_{ij} \sigma^2_j + \sum_{j,k \ne i}  c_{ijk} \sigma_j \sigma_k + ...$, where $a_{ij}, b_{ij}, c_{ijk}$ are the series coefficients, and we have omitted the constant in the expansion. Since $\sigma_j \in \{0,1\}$, $\sigma^2_j = \sigma_j$, and the $b_{ij} \sigma^2_j$ terms can be grouped with the $a_{ij} \sigma_j$ ones, this yields $x_i = h _i + \sum_{j \ne i} J_{ij} \sigma_j + \sum_{j,k \ne i}  \gamma_{ijk} \sigma_j \sigma_k + ...$, where $J_{ij} = a_{ij} + b_{ij}$ and $\gamma_{ijk} = c_{ijk}$.

\section{Analytic Calculations of Mutual Information}
\label{sec:MI_calc}
This Section is organized as follows: we first parametrize the strength of the stimulus-coupling ($\epsilon$) in the probability model. Next, we re-write the mutual information in a more convenient form. Then, we exploit that re-write to expand the MI in powers of $\epsilon$.

\subsection{Problem (re)Statement}

We want to compute the mutual information (MI) between neural responses $\vec{\sigma}$, where $\sigma_i \in \{0,1\}$ is the spiking or not spiking of a neuron in a given time bin, and the stimuli $\vec{h^s}$. We do this by writing down the conditional PDF, and computing conditional and marginal entropies. Following the main paper paper (Eq. 4), let's write down the conditional PDF as 

\begin{eqnarray}
p(\vec{\sigma}| \vec{h^s})= \frac{1}{Z(\vec{h^s})}\exp \left[\epsilon \vec{h^s}\cdot \vec{\sigma} + \beta \left(\sum_{ij} J_{ij} \sigma_i \sigma_j + \sum_{ijk} \gamma_{ijk} \sigma_i \sigma_j   \sigma_k   \right)   \right],
\end{eqnarray} 

where the pair-wise and triplet interactions are given by $J$ and $\gamma$, and $\beta$ describes the strength of those interactions. Note that the diagonal elements of $J$ can be thought of as the bias terms ($h^0_i$ in the main paper) for the neurons. As in Eq. 4 of the main paper, parameter $\epsilon$ describes the strength of the coupling between stimuli and neural responses. $Z(\vec{h^s})$ is the conditional partition function,

\begin{eqnarray}
Z( \vec{h^s})= \sum_{\{\vec{\sigma} \}} \exp \left[\epsilon \vec{h^s}\cdot \vec{\sigma}  + \beta \left( \sum_{ij} J_{ij} \sigma_i \sigma_j + \sum_{ijk} \gamma_{ijk} \sigma_i \sigma_j   \sigma_k \right)    \right].
\end{eqnarray}

The mutual information can then be computed as

\begin{eqnarray}
MI = - \sum_{ \{ \vec{\sigma} \}} p(\vec{\sigma}) \log[  p(\vec{\sigma})  ]   + \int d \vec{h^s} p(\vec{h^s}) \sum_{ \{ \vec{\sigma} \}} p(\vec{\sigma}| \vec{h^s}) \log[  p(\vec{\sigma}| \vec{h^s})  ].  
\end{eqnarray}

Note that these two terms look like $\log(p)$ averaged over some distributions, which requires that we know the partition function (i.e.: the $\log(Z(\vec{h^s})$ terms). Computing partition functions is \emph{hard}, since it requires a sum over all states, and there are $2^N$ states of a system with N neurons. The ``$\alpha$ method", described below, gets around this difficulty.

\subsection{The ``$\alpha$ method"}

\subsubsection{Re-writing the conditional PDF}

Looking at the conditional PDF, we notice that (ignoring for now the partition function), it factorizes into two parts. First, the stim-coupling part $\alpha = \exp \left[\epsilon \left( \vec{h^s}\cdot \vec{\sigma}  \right) \right]$. And, next, the ``network interactions" part $\psi = \exp \left[\beta \left( \sum_{ij} J_{ij} \sigma_i \sigma_j + \sum_{ijk} \gamma_{ijk} \sigma_i \sigma_j   \sigma_k  \right) \right]$. Consider, for now, the partition function when there is no stimulus present, so that $\vec{h^s} = \vec{0}$. In that case, $\alpha = 1$, and the partition function becomes

\begin{eqnarray}
Z( \vec{h^s} = 0)= \sum_{\vec{\sigma}} \exp \left[ \beta \left(  \sum_{ij}J_{ij} \sigma_i \sigma_j + \sum_{ijk} \gamma_{ijk} \sigma_i \sigma_j   \sigma_k \right)   \right] = \sum_{\vec{\sigma}} \psi .
\end{eqnarray}

Call this ``zero-stim" partition function $Z_0$. We can then notice that PDF of the spontaneous activity distribution (or ``zero-stim" distribution) is

\begin{eqnarray}
p(\vec{\sigma}| \vec{h^s}=0)= \frac{1}{Z_0} \exp \left[ \beta \left(\sum_{ij} J_{ij} \sigma_i \sigma_j + \sum_{ijk} \gamma_{ijk} \sigma_i \sigma_j   \sigma_k \right) \right] = \frac{\psi}{Z_0}.
\end{eqnarray} 

And, accordingly, the non-zero-field partition function is

\begin{eqnarray}
Z( \vec{h^s}) &=& \sum_{\vec{\sigma}} \exp \left[\epsilon \left( \vec{h^s}\cdot \vec{\sigma} \right) +  \beta \left( \sum_{ij} J_{ij} \sigma_i \sigma_j + \sum_{ijk} \gamma_{ijk} \sigma_i \sigma_j   \sigma_k  \right)  \right]. \\
 &=& \sum_{\vec{\sigma}}  \alpha \psi \nonumber \\
 &=& \sum_{\vec{\sigma}}  \alpha Z_0 p(\vec{\sigma}| \vec{h^s}=0) \nonumber \\
 &=& Z_0 \E_{SAD} \left[ \alpha \right], \nonumber
\end{eqnarray}

where $ \E_{SAD} \left[ \cdot \right]$ represents an expectation over the spontaneous activity distribution (SAD). Finally, using this bit of notation, we can re-write our conditional PDF as 

\begin{eqnarray}
p(\vec{\sigma}| \vec{h^s}) &=& \frac{1}{Z(\vec{h^s})} \alpha \psi,  \\
&=& \frac{\alpha \psi}{Z_0 \E_{SAD} \left[ \alpha \right]}  \nonumber \\
&=& \frac{\alpha p(\vec{\sigma}| \vec{h^s}=0) }{\E_{SAD} \left[ \alpha \right]}. \nonumber
\end{eqnarray} 

This, on its own, doesn't look like much of a simplification. However, notice that we no longer have a conditional partition function to evaluate. In its stead, we have the zero-field partition function, and an expectation of $\alpha$ over the SAD. The zero-field partition function will be the same for the conditional and marginal distributions, and will thus cancel when we compute mutual information (below).

\subsubsection{Simplifying the Mutual Information}

We use another form for the mutual information, that is equivalent to (Eq. B3), but a bit more convenient for our purposes:

\begin{eqnarray}
MI = \int d \vec{h^s} p(\vec{h^s}) \sum_{ \{ \vec{\sigma} \}} p(\vec{\sigma}| \vec{h^s}) \log[  p(\vec{\sigma}| \vec{h^s}) / p(\vec{\sigma})  ].  
\end{eqnarray}

To evaluate this, we need to know the marginal distribution over responses, which is

\begin{eqnarray}
p(\vec{\sigma}) &=&  \int d \vec{h^s} p(\vec{h^s})  p(\vec{\sigma}| \vec{h^s}) \\
&=&   p(\vec{\sigma}| \vec{h^s}=0)  \E_{\vec{h^s}} \left[ \frac{\alpha  }{\E_{SAD} \left[ \alpha \right] } \right] , \nonumber
\end{eqnarray} 

where $ \E_{\vec{h^s}} \left[\cdot \right] $ is an expectation over the stimulus distribution. Using all of our above observations, we can re-write the mutual information as

\begin{eqnarray}
MI &=&\E_{\vec{h^s}}\left[  \sum_{ \{ \vec{\sigma} \}} p(\vec{\sigma}| \vec{h^s}=0) \frac{\alpha }{ \E_{SAD} \left[ \alpha \right] } \log \left( \frac{ \frac{\alpha }{ \E_{SAD} \left[ \alpha \right] } }{   \E_{\vec{h^s}} \left[ \frac{\alpha  }{ \E_{SAD}\left[ \alpha \right]} \right] } \right)   \right] \\
&=&  \E_{\vec{h^s}}\left[  \E_{SAD} \left[ \frac{\alpha }{ \E_{SAD} \left[ \alpha \right] } \log \left( \frac{ \frac{\alpha }{ \E_{SAD} \left[ \alpha \right] } }{   \E_{\vec{h^s}} \left[ \frac{\alpha  }{ \E_{SAD}\left[ \alpha \right]} \right] } \right)   \right] \right] \nonumber
\end{eqnarray}

where the factors of $p(\vec{\sigma}| \vec{h^s}=0) $ in the logarithm cancel because they appear in both the numerator and denominator of the fraction, and the second line follows from the first because the sum over all states reduces to an average over the spontaneous activity distribution. We now observe that the problem of computing mutual information, which previously required a \emph{hard} partition function calculation, has been reduced to computing expectations of $\alpha = \exp(\epsilon \vec{h^s} \cdot \vec{\sigma})$ over both the spontaneous activity distribution, and over the stimulus distribution. 

Finally, notice that, while we have used a specific functional form for $p(\vec{\sigma} | \vec{h^s})$ in our calculations, the same logic applies to any conditional PDF of the form $p(\vec{\sigma} | \vec{h^s}) = \alpha(\vec{h^s},\vec{\sigma}) \psi(\vec{\sigma})$, with the condition $\alpha(\vec{h^s} = 0, \vec{\sigma}) \ne 0$. Intuitively, the variability in responses due to $\psi(\vec{\sigma})$ is the same for all stimuli, so it provides the same contribution to the marginal and conditional entropies, and that contribution cancels when we compute mutual information.

\subsection{Analytically relating MI to moments of the stimulus and response distributions}

Our MI calculation reduced to (Eq. B10) expectations over the SAD and the stimulus distribution. We know that we can write out the exponentials, logarithms, and ratios as power series in $\epsilon \vec{h^s} \cdot \vec{\sigma}$, and the MI is an average of those things over the SAD and stimulus distributions. This means that we can analytically write down the MI as a function of moments of the SAD and stimulus distributions. Let's do that, but first break up Eq. B10 into a few easier-to-manage terms:

\begin{eqnarray}
MI &=&  \E_{\vec{h^s}}\left[  \E_{SAD} \left[ \frac{\alpha }{ \E_{SAD} \left[ \alpha \right] } \log \left( \frac{ \frac{\alpha }{ \E_{SAD} \left[ \alpha \right] } }{   \E_{\vec{h^s}} \left[ \frac{\alpha  }{ \E_{SAD}\left[ \alpha \right]} \right] } \right)   \right] \right] \\
  &=&   \E_{\vec{h^s}}\left[  \E_{SAD} \left[ \log\left( \alpha \right)  \frac{\alpha }{ \E_{SAD} \left[ \alpha \right] }     \right] \right] -   \E_{\vec{h^s}}\left[  \log \left(   \E_{SAD}\left[ \alpha \right]  \right) \right]  \nonumber \\
  &-&  \E_{\vec{h^s}}\left[  \E_{SAD} \left[ \frac{\alpha }{ \E_{SAD} \left[ \alpha \right] } \log \left(  \E_{\vec{h^s}} \left[ \frac{\alpha  }{ \E_{SAD}\left[ \alpha \right]} \right]  \right)   \right] \right]  \nonumber
\end{eqnarray}

\subsubsection{Dealing with the $3^{rd}$ term in the above equation}

The first few terms are (relatively) simple, so let's focus for now on the third term ($-T_3$) in the above equation, with:

\begin{eqnarray}
T_3= \E_{\vec{h^s}}\left[  \E_{SAD} \left[ \frac{\alpha }{ \E_{SAD} \left[ \alpha \right] } \log \left(  \E_{\vec{h^s}} \left[ \frac{\alpha  }{ \E_{SAD}\left[ \alpha \right]} \right]  \right)   \right] \right] .
\end{eqnarray}

We'll expand to $3^{rd}$ order in $\epsilon$, and thus restrict ourselves to small $\epsilon$. In so doing, we will find that $T_3 = 0 + \mathcal{O}\left(\epsilon^4\right)$, and hence that we can ignore it in the small $\epsilon$ limit. To simplify our notation, let $s = \vec{\sigma} \cdot \vec{h^s}$ (the $s$ is for ``sum"), such that $\alpha = \exp(\epsilon s)$. We first note that $T_3(0) = 0$, because $\alpha(\epsilon = 0) = 1~\forall~s$, and thus the ratio in the logarithm is unity. Since they will keep coming up in our calculation, we'll note straight away that

\begin{eqnarray}
&\frac{\partial}{\partial \epsilon} \left( \frac{\alpha}{ \E_{SAD}[\alpha]} \right)= \frac{s \alpha E_{SAD}[\alpha] -  \alpha E_{SAD}[s\alpha] }{ \left( E_{SAD}[\alpha] \right)^2} \\
&\frac{\partial^2}{\partial \epsilon^2} \left( \frac{\alpha}{ \E_{SAD}[\alpha]} \right) = \Omega \nonumber \\
&\Omega = \frac{s^2 \alpha \left(\E_{SAD}[\alpha]\right)^2 - \alpha \E_{SAD}[ s^2 \alpha]  \E_{SAD}[ \alpha]  - 2 s\alpha \E_{SAD}[ s \alpha]  \E_{SAD}[ \alpha]  +  2 \alpha  \left(\E_{SAD}[ s \alpha]\right)^2 }{ \E_{SAD}^3 \left[\alpha \right]   } ,     \nonumber 
\end{eqnarray}

where we have defined $\Omega$ for our convenience. Then,

\begin{eqnarray}
T_3 (\epsilon) = T_3(0) + \epsilon \frac{\partial T_3}{\partial \epsilon} \bigg|_{\epsilon=0} + \frac{\epsilon^2}{2} \frac{\partial^2 T_3}{\partial \epsilon^2} \bigg|_{\epsilon=0} + \frac{\epsilon^3}{6} \frac{\partial^3 T_3}{\partial \epsilon^3} \bigg|_{\epsilon=0} + \mathcal{O}(\epsilon^4).
\end{eqnarray}

To proceed further, we need to evaluate the derivatives in Eq. B14.

\begin{eqnarray}
\frac{\partial T_3}{\partial \epsilon} &=&   \E_{\vec{h^s}} \left[ \E_{SAD} \left[   \frac{\alpha  }{ \E_{SAD}}  \left( \E_{\vec{h^s}} \left[  \frac{\alpha  }{ \E_{SAD}} \right] \right)^{-1}       \E_{\vec{h^s}} \left[  \frac{s \alpha \E_{SAD}[\alpha] - \alpha \E_{SAD} [s \alpha]  }{ \E_{SAD}^2 \left[\alpha \right]   }                           \right]          \right] \right] \\
&+&  \E_{\vec{h^s}} \left[ \E_{SAD} \left[  \log \left(  \E_{\vec{h^s}} \left[ \frac{\alpha  }{ \E_{SAD}\left[ \alpha \right]} \right]  \right) \frac{s \alpha \E_{SAD}[\alpha] - \alpha \E_{SAD} [s \alpha]  }{ \E_{SAD}^2 \left[\alpha \right]   }   \right] \right]. \nonumber
\end{eqnarray}

Conveniently, the first term in Eq. B15 (above) vanishes. To see this, notice that $ \left( \E_{\vec{h^s}} \left[  \frac{\alpha  }{ \E_{SAD}} \right] \right)^{-1} $ and $   \E_{\vec{h^s}} \left[  \frac{s \alpha \E_{SAD}[\alpha] - \alpha \E_{SAD} s \alpha]  }{ \E_{SAD}^2 \left[\alpha \right]   }                           \right]   $ are independent of $\vec{h^s}$, being already expectation values over $\vec{h^s}$. If we change the order of our expectations over SAD and $\vec{h^s}$, then $ \frac{\alpha  }{ \E_{SAD}}$ and $ \left( \E_{\vec{h^s}} \left[  \frac{\alpha  }{ \E_{SAD}} \right] \right)^{-1} $ cancel. What remains is (again, swapping the order in which we compute expectations)

\begin{eqnarray}
&&\E_{\vec{h^s}} \left[ \E_{SAD} \left[  \frac{s \alpha \E_{SAD}[\alpha] - \alpha \E_{SAD} [s \alpha]  }{ \E_{SAD}^2 \left[\alpha \right]   }                           \right]          \right]  \\
&=&  \E_{\vec{h^s}}  \left[  \frac{ \E_{SAD}[s \alpha ] \E_{SAD}[\alpha] - \E_{SAD}[ \alpha] \E_{SAD} [s \alpha]  }{ \E_{SAD}^2 \left[\alpha \right]   }                           \right]          \nonumber \\
&=& 0 \nonumber.
\end{eqnarray}

With this simplification, we have

\begin{eqnarray}
\frac{\partial T_3}{\partial \epsilon} = \E_{SAD} \left[  \log \left(  \E_{\vec{h^s}} \left[ \frac{\alpha  }{ \E_{SAD}\left[ \alpha \right]} \right]  \right)  \E_{\vec{h^s}} \left[ \frac{s \alpha \E_{SAD}[\alpha] - \alpha \E_{SAD} [s \alpha]  }{ \E_{SAD}^2 \left[\alpha \right]   }   \right] \right],
\end{eqnarray}  

which is zero when evaluated at $\epsilon = 0$, due to the ratio of $\alpha$'s in the logarithm. There is thus no first-order contribution to $T_3$! We now require the second (and eventually third) derivative in our Taylor series:

\begin{eqnarray}
\frac{\partial^2 T_3}{\partial \epsilon^2} &=&   \E_{SAD} \left[  \log \left(  \E_{\vec{h^s}} \left[ \frac{\alpha  }{ \E_{SAD}\left[ \alpha \right]} \right]  \right) \E_{\vec{h^s}} \left[ \Omega \right] \right] + \zeta \\
\zeta &=&  \E_{SAD} \left[ \left(  \E_{\vec{h^s}} \left[ \frac{\alpha  }{ \E_{SAD}\left[ \alpha \right]} \right]  \right)^{-1} \left( \E_{\vec{h^s}} \left[ \frac{s \alpha \E_{SAD}[\alpha] - \alpha \E_{SAD} [ s \alpha]  }{ \E_{SAD}^2 \left[\alpha \right]   }   \right] \right)^2 \right] . \nonumber
\end{eqnarray}  

As with the first derivative, when we evaluate this at $\epsilon = 0$, the term with the logarithm vanishes, leaving

\begin{eqnarray}
 \frac{\partial^2 T_3}{\partial \epsilon^2} \bigg|_{\epsilon=0} &=& \zeta \bigg|_{\epsilon=0} \\
 &=&   \E_{SAD} \left[      \left(  \E_{\vec{h^s}}[s] -  \E_{\vec{h^s}} \left[ \E_{SAD}[s]  \right] \right)^2        \right]   \nonumber \\
 \implies   \frac{\partial^2 T_3}{\partial \epsilon^2} \bigg|_{\epsilon=0} &=& \E_{SAD}  \left[  \left( \E_{\vec{h^s}}[s] \right)^2 \right] - \left(  \E_{SAD,\vec{h^s}} [s]    \right)^2 \nonumber
 \end{eqnarray}

The second derivative was messy to calculate (Eq. B18), and now we require the \emph{third} derivative, which is even worse. We'll save ourselves a bit of effort by noting that the term that contains a logarithm will vanish when evaluated at $\epsilon =  0$, and so we won't compute it at all. With that in mind,

\begin{eqnarray}
\frac{\partial^3 T_3}{\partial \epsilon^3} &=&  \E_{SAD} \left[  \log \left(  \E_{\vec{h^s}} \left[ \frac{\alpha  }{ \E_{SAD}\left[ \alpha \right]} \right]  \right)   \E_{\vec{h^s}} \left[  \frac{\partial \Omega}{\partial \epsilon } \right] \right]  \\
&+&  \E_{SAD} \left[ \left(  \E_{\vec{h^s}} \left[ \frac{\alpha  }{ \E_{SAD}\left[ \alpha \right]} \right]  \right)^{-1}  \left( \E_{\vec{h^s}} \left[ \frac{s \alpha \E_{SAD}[\alpha] - \alpha \E_{SAD} [s \alpha]  }{ \E_{SAD}^2 \left[\alpha \right]   }   \right] \right)  \E_{\vec{h^s}} \left[ \Omega \right] \right] \nonumber \\
&+& \frac{\partial \zeta}{\partial \epsilon} \nonumber,
\end{eqnarray}  

where $\Omega$ and $\zeta$ are as defined in Eqs. B13, B18, and

\begin{eqnarray}
\frac{\partial \zeta}{\partial \epsilon} &=&   2 \E_{SAD} \left[ \left(  \E_{\vec{h^s}} \left[ \frac{\alpha  }{ \E_{SAD}\left[ \alpha \right]} \right]  \right)^{-1} \left( \E_{\vec{h^s}} \left[ \frac{s \alpha \E_{SAD}[\alpha] - \alpha \E_{SAD} [s \alpha]  }{ \E_{SAD}^2 \left[\alpha \right]   }   \right] \right) \E_{\vec{h^s}}\left[ \Omega \right] \right] \\
&-&  \E_{SAD} \left[ \left(  \E_{\vec{h^s}} \left[ \frac{\alpha  }{ \E_{SAD}\left[ \alpha \right]} \right]  \right)^{-2} \left( \E_{\vec{h^s}} \left[ \frac{s \alpha \E_{SAD}[\alpha] - \alpha \E_{SAD} [s \alpha]  }{ \E_{SAD}^2 \left[\alpha \right]   }   \right] \right)^3 \right]. \nonumber
\end{eqnarray}  

Putting all of these pieces together, we observe that 

\begin{eqnarray}
\frac{\partial^3 T_3}{\partial \epsilon^3}  &=&  \E_{SAD} \left[  \log \left(  \E_{\vec{h^s}} \left[ \frac{\alpha  }{ \E_{SAD}\left[ \alpha \right]} \right]  \right)    \E_{\vec{h^s}} \left[ \frac{\partial \Omega}{\partial \epsilon } \right] \right]  \\
&+& 3 \E_{SAD} \left[ \left(  \E_{\vec{h^s}} \left[ \frac{\alpha  }{ \E_{SAD}\left[ \alpha \right]} \right]  \right)^{-1} \left( \E_{\vec{h^s}} \left[ \frac{s \alpha \E_{SAD}[\alpha] - \alpha \E_{SAD} [s \alpha]  }{ \E_{SAD}^2 \left[\alpha \right]   }   \right] \right) \E_{\vec{h^s}} \left[ \Omega \right] \right]  \nonumber \\
&-&  \E_{SAD} \left[ \left(  \E_{\vec{h^s}} \left[ \frac{\alpha  }{ \E_{SAD}\left[ \alpha \right]} \right]  \right)^{-2} \left( \E_{\vec{h^s}} \left[ \frac{s \alpha \E_{SAD}[\alpha] - \alpha \E_{SAD} [s \alpha]  }{ \E_{SAD}^2 \left[\alpha \right]   }   \right] \right)^3 \right], \nonumber
\end{eqnarray}

where (again) $\Omega$ is as defined in Eq. B13. Let's now evaluate this at $\epsilon = 0$;

\begin{eqnarray}
\frac{\partial^3 T_3}{\partial \epsilon^3}  \bigg|_{\epsilon=0} &=& 3 \E_{SAD} \left[ \left( \E_{\vec{h^s}} \left[ s \right] - \E_{SAD,\vec{h^s}} [s]    \right) \E_{\vec{h^s}} \left [ \Omega  \bigg|_{\epsilon=0} \right] \right] \nonumber  \\
&-&  \E_{SAD} \left[ \left( \E_{\vec{h^s}} \left[s \right] -  \E_{SAD,\vec{h^s}} [ s ]  \right)^3 \right].
\end{eqnarray}

This looks pretty hairy, but we'll note that all of our terms are multiplied by either $\E_{\vec{h^s}}(s)$ or  $\E_{SAD,\vec{h^s}}(s)$. These will vanish for zero-mean stimulus distributions. Since one can compensate for non-zero mean along any stimulus distribution by changing the bias for the corresponding neuron (the diagonal elements $J_{ii}$ in Eq. B2), then WLOG we can assume the stimuli are zero-mean. This means that

\begin{eqnarray}
\E_{\vec{h^s}} [s] &=& \E_{\vec{h^s}} \left[ \sum_i h_i \sigma_i \right] \\
&=&   \sum_i  \E_{\vec{h^s}} \left[ h_i \right ]  \sigma_i  \nonumber \\
&=& 0 ~\forall ~\vec{\sigma}. \nonumber
\end{eqnarray}

Thus, the $3^{rd}$ order term vanishes Looking back at Eq. B19, one can similarly note that the second order term vanishes, and (at least up to $3^{rd}$ order in $\epsilon$), $T_3$ vanishes!

\subsubsection{Looking back at the other terms in our MI formula}

Noting (from above) that we can ignore $T_3$  up to $3^{rd}$ order in $\epsilon$, we find that

\begin{eqnarray}
MI &=&  \E_{\vec{h^s}}\left[  \E_{SAD} \left[ \log\left( \alpha \right)  \frac{\alpha }{ \E_{SAD} \left[ \alpha \right] }     \right] \right] -   \E_{\vec{h^s}}\left[  \log \left(   \E_{SAD}\left[ \alpha \right]  \right) \right]  + \mathcal{O}(\epsilon^4),
  \end{eqnarray}

where we must now evaluate the two terms in Eq. B25 in order to proceed. Using the fact that $\alpha = \exp(\epsilon s)$, and calling these terms $T_1$ and $T_2$,

\begin{eqnarray}
T_1 &=&  \E_{\vec{h^s}}\left[  \E_{SAD} \left[\epsilon s  \frac{\alpha }{ \E_{SAD} \left[ \alpha \right] }     \right] \right] \\
T_2 &=& \E_{\vec{h^s}}\left[  \log \left(   \E_{SAD}\left[ \alpha \right]  \right) \right]  \nonumber.
  \end{eqnarray}

\subsubsection{Expanding the first term, $T_1$}

We'll start by expanding the $T_1$ term up to $3^{rd}$ order in $\epsilon$, but first note that it already has a factor of $\epsilon$ in front of it (from the $\log(\alpha)$), so we will need do a second-order Taylor series of $\tilde{T_1} = T_1/\epsilon$, and just multiply that whole thing by $\epsilon$. Thus,

\begin{eqnarray}
T_1(\epsilon) &=& \epsilon \left[  \tilde{T_1}(\epsilon)         \right] \\
&=&  \epsilon \left[  \tilde{T_1}(\epsilon = 0) + \epsilon \frac{\partial \tilde{T_1}}{\partial \epsilon}  \bigg|_{\epsilon=0}  + \frac{\epsilon}{2} \frac{\partial^2 \tilde{T_1}}{\partial \epsilon^2}  \bigg|_{\epsilon=0}           \right] + \mathcal{O}(\epsilon^4). \nonumber
\end{eqnarray}

The $0^{th}$ order term in the expansion vanishes  ($ \tilde{T_1}(\epsilon = 0) = \E_{SAD,\vec{h^s}}[s] = 0$) for the same reason our $T_3$ vanished (Eq. B25). Again, we require derivatives to proceed, and (Eq. B13)

\begin{eqnarray}
\frac{\partial \tilde{T_1}}{\partial \epsilon} =  \E_{SAD,\vec{h^s}} \left[  s \frac{s \alpha E_{SAD}[\alpha] - \alpha E_{SAD}[s \alpha]   }{\left(E_{SAD}[\alpha]\right)^2}         \right].
\end{eqnarray}

Evaluating this at $\epsilon = 0$, we observe that

\begin{eqnarray}
\frac{\partial \tilde{T_1}}{\partial \epsilon} \bigg|_{\epsilon=0} &=&  \E_{SAD,\vec{h^s}} \left[  s \left( s -  E_{SAD}[s]  \right)  \right] \\
&=&  \E_{SAD,\vec{h^s}} \left[ s^2 \right] - E_{\vec{h^s}}\left[    \left( E_{SAD}[s] \right)^2     \right] , \nonumber
\end{eqnarray}

which does not necessarily vanish for zero-mean stimulus distributions! Carrying on to get the second derivative,

\begin{eqnarray}
\frac{\partial^2 \tilde{T_1}}{\partial \epsilon^2} =  \E_{SAD,\vec{h^s}} \left[  s \Omega        \right],
\end{eqnarray}

where $\Omega$ (Eq. B13) has made a re-appearance. Evaluating $\Omega$ for $\epsilon = 0$, we find that
\begin{eqnarray}
\frac{\partial^2 \tilde{T_1}}{\partial \epsilon^2}  \bigg|_{\epsilon=0} &=& \E_{SAD,\vec{h^s}} \left[  s \left( s^2 - \E_{SAD}[s^2] - 2s \E_{SAD}[s] + 2 (\E_{SAD}[s])^2        \right) \right] \\
&=&  \E_{SAD,\vec{h^s}} \left[s^3 \right] -  \E_{SAD,\vec{h^s}} \left[ s  \E_{SAD} \left[ s^2 \right] \right]  - 2   \E_{SAD,\vec{h^s}} \left[ s^2  \E_{SAD} \left[ s \right] \right]  + 2   \E_{SAD,\vec{h^s}} \left[ s \left(  \E_{SAD} \left[ s \right]  \right)^2 \right]. \nonumber
\end{eqnarray}

Thus, combining Eqs. B27, B29 and B31, $T_1$ (Eq. B26) is

\begin{eqnarray}
T_1 &=& \epsilon^2 \left( \E_{SAD,\vec{h^s}} \left[ s^2 \right] - E_{\vec{h^s}}\left[    \left( E_{SAD}[s] \right)^2     \right]  \right) \\
&+& \frac{\epsilon^3}{2} \left(  \E_{SAD,\vec{h^s}} \left[s^3 \right] -  \E_{SAD,\vec{h^s}} \left[ s  \E_{SAD} \left[ s^2 \right] \right]  - 2   \E_{SAD,\vec{h^s}} \left[ s^2  \E_{SAD} \left[ s \right] \right]  + 2   \E_{SAD,\vec{h^s}} \left[ s \left(  \E_{SAD} \left[ s \right]  \right)^2 \right] \right). \nonumber
\end{eqnarray}

So, the first term in our MI function has been reduced to moments of the stimulus and spontaneous activity distributions, which we will later simplify.

\subsubsection{Expanding the second term, $T_2$}

Let's go over our second term now ($T_2$ in Eq. B26):

\begin{eqnarray}
T_2 (\epsilon) &=&  \E_{\vec{h^s}}\left[  \log \left(   \E_{SAD}\left[ \alpha \right]  \right) \right]  \\
&=&T_2(0) + \epsilon \frac{\partial T_2}{\partial \epsilon} \bigg|_{\epsilon=0} + \frac{\epsilon^2}{2} \frac{\partial^2 T_2}{\partial \epsilon^2} \bigg|_{\epsilon=0} + \frac{\epsilon^3}{6} \frac{\partial^3 T_2}{\partial \epsilon^3} \bigg|_{\epsilon=0} + \mathcal{O}(\epsilon^4). \nonumber
\end{eqnarray}

The $0^{th}$ order term $T_2(0)$ vanishes due to the log. Let's compute the derivatives one-by-one, starting with the first derivative:

\begin{eqnarray}
\frac{\partial T_2}{\partial \epsilon}  = \E_{\vec{h^s}}\left[   \frac{\E_{SAD}\left[ s \alpha \right]}{\E_{SAD}\left[ \alpha \right] }  \right].
\end{eqnarray}

When evaluated at $\epsilon = 0$, this derivative vanishes for zero-mean stimulus distributions:

\begin{eqnarray}
\frac{\partial T_2}{\partial \epsilon}  \bigg|_{\epsilon=0}  = \E_{\vec{h^s}}\left[  \E_{SAD}\left[ s \right]  \right] = 0,
\end{eqnarray}

where, in the second step, we swap the order in which we take the averages, and recall Eq. B24. Moving on to the second derivative,

\begin{eqnarray}
\frac{\partial^2 T_2}{\partial \epsilon^2}  = \E_{\vec{h^s}}\left[   \frac{  \E_{SAD}\left[ \alpha \right] \E_{SAD}\left[ s^2 \alpha \right] - \E^2_{SAD}\left[ s \alpha \right]}{\E^2_{SAD}\left[ \alpha \right] }  \right].
\end{eqnarray}

When evaluated at $\epsilon = 0$, we find that

\begin{eqnarray}
\frac{\partial^2 T_2}{\partial \epsilon^2}  \bigg|_{\epsilon=0}   = \E_{\vec{h^s}}\left[    \E_{SAD}\left[ s^2 \right] - \left(\E_{SAD}\left[ s \right] \right)^2   \right].
\end{eqnarray}

Finally, we require the third derivative,

\begin{eqnarray}
\frac{\partial^3 T_2}{\partial \epsilon^3}  &=& \E_{\vec{h^s}}\left[   \frac{  -2\E_{SAD}[\alpha] \E_{SAD}[s\alpha] \left( \E_{SAD}\left[ \alpha \right] \E_{SAD}\left[ s^2 \alpha \right] - \E^2_{SAD}\left[ s \alpha \right] \right)            }{\E^4_{SAD}\left[ \alpha \right] }  \right] \\
&+&  \E_{\vec{h^s}}\left[   \frac{  \E^2_{SAD}[\alpha] \left(- \E_{SAD}\left[ s  \alpha \right] \E_{SAD}\left[ s^2 \alpha \right] + \E_{SAD}\left[ \alpha \right] \E_{SAD}\left[ s^3 \alpha \right]    \right)            }{\E^4_{SAD}\left[ \alpha \right] }  \right],  \nonumber
\end{eqnarray}

which could be simplified, but will be easier to simplify once we evaluate it at $\epsilon = 0$:

\begin{eqnarray}
\frac{\partial^3 T_2}{\partial \epsilon^3}  \bigg|_{\epsilon=0}   = \E_{\vec{h^s}}\left[  2 (\E_{SAD}[s])^3  -3  \E_{SAD}[s] \E_{SAD}[s^2]  +  \E_{SAD}[s^3]    \right].
\end{eqnarray}

Assembling the pieces (Eqs. B33,B35,B37,B39), we find that 

\begin{eqnarray}
T_2 &=& \frac{\epsilon^2}{2} \E_{\vec{h^s}}\left[    \E_{SAD}\left[ s^2 \right] - \left(\E_{SAD}\left[ s \right] \right)^2   \right] \\
&+& \frac{\epsilon^3}{6}\E_{\vec{h^s}}\left[  2 (\E_{SAD}[s])^3  -3  \E_{SAD}[s] \E_{SAD}[s^2]  +  \E_{SAD}[s^3]    \right] + \mathcal{O}(\epsilon^4). \nonumber
\end{eqnarray}

\subsubsection{Assembling to terms in the MI}

We can now take the individual terms we calculated (Eqs. B32,B40 for terms $T_1$ and $T_2$), and assemble them to find the MI (Eq. B25) 

\begin{eqnarray}
MI &=& \epsilon^2 \left( \E_{SAD,\vec{h^s}} \left[ s^2 \right] - E_{\vec{h^s}}\left[    \left( E_{SAD}[s] \right)^2     \right]  \right) \\
&+& \frac{\epsilon^3}{2} \left(  \E_{SAD,\vec{h^s}} \left[s^3 \right] -  \E_{SAD,\vec{h^s}} \left[ s  \E_{SAD} \left[ s^2 \right] \right]  - 2   \E_{SAD,\vec{h^s}} \left[ s^2  \E_{SAD} \left[ s \right] \right]  + 2   \E_{SAD,\vec{h^s}} \left[ s \left(  \E_{SAD} \left[ s \right]  \right)^2 \right] \right). \nonumber \\
&-& \frac{\epsilon^2}{2} \E_{\vec{h^s}}\left[    \E_{SAD}\left[ s^2 \right] - \left(\E_{SAD}\left[ s \right] \right)^2   \right] \nonumber \\
&-& \frac{\epsilon^3}{6}\E_{\vec{h^s}}\left[  2 (\E_{SAD}[s])^3  -3  \E_{SAD}[s] \E_{SAD}[s^2]  +  \E_{SAD}[s^3]    \right] + \mathcal{O}(\epsilon^4). \nonumber
\end{eqnarray}

Grouping the $\epsilon^2$ and $\epsilon^3$ terms together, 

\begin{eqnarray}
MI &=& \frac{\epsilon^2}{2} \left( \E_{SAD,\vec{h^s}} \left[ s^2 \right] - E_{\vec{h^s}}\left[    \left( E_{SAD}[s] \right)^2     \right]  \right) \\
&+& \frac{\epsilon^3}{3} \left(  \E_{SAD,\vec{h^s}} \left[s^3 \right]  - \E_{\vec{h^s}}\left[   (\E_{SAD}[s])^3 \right] \right) \nonumber \\ 
&+& \epsilon^3 \left(    \E_{\vec{h^s}} \left[ \left(  \E_{SAD} \left[ s \right]  \right)^3 \right]   -      \E_{\vec{h^s}} \left[  \E_{SAD}[s^2]  \E_{SAD} \left[ s \right] \right]       \right)+  \mathcal{O}(\epsilon^4). \nonumber
\end{eqnarray}

\subsubsection{Computing the relevant moments of the SAD and stimulus distributions}

To proceed, we need to compute the expectation values in Eq. B42. Let's start with the $\epsilon^2$ terms, and notice that

\begin{eqnarray}
s^2 &=& \sum_{ij} \sigma_i \sigma_j h_i h_j \\
&=& \sum_i \sigma_i h^2_i + \sum_{i\ne j} \sigma_i \sigma_j h_i h_j, \nonumber
\end{eqnarray}

where we have used the fact that, for $\sigma_i \in\{0,1\}$, $\sigma_i^2 = \sigma_i$. Averaging these over the SAD, with $\E_{SAD}[\sigma_i] = \mu_i$, $\E_{SAD}[\sigma_i \sigma_j] = \pi_{ij}$ (the $\pi$ is for ``pair"), we see that

\begin{eqnarray}
\E_{SAD} [s^2]  = \sum_i \mu_i h^2_i + \sum_{i\ne j} \pi_{ij} h_i h_j.
\end{eqnarray}

Averaging this quantity over the stimulus distribution, recalling that the stimuli are zero-mean, and denoting the elements of the covariance matrix by $\nu_{ij} = \E_{\vec{h^s}}[h_i h_j]$ (we can't use $\sigma_{ij}$ to denote covariance since that already describes the spin states),

\begin{eqnarray}
\E_{SAD,\vec{h^s}} [s^2]  = \sum_i \mu_i \nu_{ii} + \sum_{i\ne j} \pi_{ij} \nu_{ij}.
\end{eqnarray}

Similarly, we can note that $ E_{\vec{h^s}}\left[    \left( E_{SAD}[s] \right)^2 \right] =  E_{\vec{h^s}}\left[  \sum_i \mu^2_i h^2_i + \sum_{ij} \mu_i \mu_j h_i h_j    \right]$. Carrying out the average over the stim distribution, we find that 

\begin{eqnarray}
 E_{\vec{h^s}}\left[    \left( E_{SAD}[s] \right)^2 \right] = \sum_i \mu^2_i \nu_{ii} + \sum_{i\ne j} \mu_i \mu_j \nu_{ij}.
\end{eqnarray}

Then, the $\epsilon^2$ term in Eq. B42 is

\begin{eqnarray}
&\frac{\epsilon^2}{2} \left[        \sum_i \mu_i \nu_{ii} + \sum_{i\ne j} \pi_{ij} \nu_{ij} -   \sum_i \mu^2_i \nu_{ii} - \sum_{i\ne j} \mu_i \mu_j \nu_{ij}  \right] \\
&= \frac{\epsilon^2}{2} \left[     \sum_i  \nu_{ii} \mu_i(1-\mu_i) + \sum_{i\ne j}  \nu_{ij} \left( \pi_{ij} - \mu_i \mu_j   \right) \right] . \nonumber
\end{eqnarray}

The first of these terms contains $ \mu_i(1-\mu_i) = \text{var}(\sigma_i)$, and the second one contains $(\pi_{ij} - \mu_i \mu_j ) = \text{cov}(\sigma_i,\sigma_j)$, where these moments (recall) are computed over the spontaneous activity distribution, and thus $\text{cov}(\sigma_i,\sigma_j)$ is best related to the $J_{ij}$ terms in Eq. B2, and $ \text{var}(\sigma_i)$ is best related to the mean-determining bias term $J_{ii}$ in Eq. B2. If we ignore the higher-order terms in the MI (the $3^{rd}$ moments come in at $\mathcal{O}(\epsilon^3)$), and restrict ourselves momentarily to $\mathcal{O}(\epsilon^2)$, we notice that

\begin{eqnarray}
MI =  \frac{\epsilon^2}{2} \left[     \sum_i  \nu_{ii} \text{var}(\sigma_i) + \sum_{i\ne j}  \nu_{ij} \text{cov}(\sigma_i,\sigma_i) \right]  + \mathcal{O}(\epsilon^3).
\end{eqnarray}

We now make our first observation, namely, that for maximum MI, we require that $\nu_{ij}$ and $\text{cov}(\sigma_i,\sigma_i)$ should have the same sign (both positive or both negative). Since the correlations in the SAD are controlled by (and have the same sign as) $J_{ij}$, we find that $J_{ij}$ should have the same sign as $\nu_{ij}$ (the stimulus covariances), which agrees with our numerical experiments, and those of Tkacik et al. We also notice that, to maximize $\text{cov}(\sigma_i,\sigma_i)$ and $\text{var}(\sigma_i)$, one should choose $\mu_i = 1/2$, which again agrees with our numerical experiments with unskewed stimuli. If there are skewed stimuli, and thus the $3^{rd}$ order terms are to be considered, these conclusions may no longer hold, so to make more progress, we must consider the $\epsilon^3$ terms in Eq. B42.

As with the $\epsilon^2$ terms, we will consider the $\epsilon^3$ terms one-by-one, starting with

\begin{eqnarray}
 \E_{SAD,\vec{h^s}} \left[s^3 \right] &=&  \E_{SAD,\vec{h^s}} \left[\sum_{ijk} \sigma_i \sigma_j \sigma_k h_i h_j h_k \right]  \\
 &=&  \E_{\vec{h^s}} \left[\sum_{i} \mu_i  h^3_i + 3 \sum_{i\ne j} \pi_{ij} h^2_i h_j + \sum_{i \ne j \ne k} \tau_{ijk} h_i h_j h_k \right] \nonumber,
\end{eqnarray}

where we have used the fact that, for $\sigma_i \in\{0,1\}$, $\sigma_i^3 = \sigma^2_i = \sigma_i$, and defined $\tau_{ijk} = \E_{SAD}[\sigma_i \sigma_j \sigma_k]$ (the $\tau$ is for ``triplet") in going from the first line to the second. Carrying out the average over the stimulus distribution, we observe that 

\begin{eqnarray}
 \E_{SAD,\vec{h^s}} \left[s^3 \right] = \sum_{i} \mu_i \left< h^3_i \right> + 3 \sum_{i\ne j} \pi_{ij} \left< h^2_i h_j  \right> + \sum_{i \ne j \ne k} \tau_{ijk} \left<h_i h_j h_k \right>,
\end{eqnarray}

where we have used triangle brackets $\left< \cdot \right>$ to denote an expectation value over the stimulus distribution. Moving on to the next $\epsilon^3$ term in Eq. B42, 

\begin{eqnarray}
\E_{\vec{h^s}}\left[   (\E_{SAD}[s])^3 \right] &=& \E_{\vec{h^s}} \left[ \sum_{ijk} \mu_i \mu_j \mu_k h_i h_j h_k  \right] \\
&=& \E_{\vec{h^s}} \left[ \sum_{i} \mu^3_i  h^3_i + 3 \sum_{i\ne j} \mu^2_i \mu_j h^2_i h_j + \sum_{i \ne j \ne k} \mu_i \mu_j \mu_k h_i h_j h_k    \right] \nonumber \\
&=&   \sum_{i} \mu^3_i  \left< h^3_i \right> + 3 \sum_{i\ne j} \mu^2_i \mu_j \left<h^2_i h_j \right> + \sum_{i \ne j \ne k} \mu_i \mu_j \mu_k \left< h_i h_j h_k\right>.\nonumber
\end{eqnarray}

We can assemble these two terms, to get the $\frac{\epsilon^3}{3}$ term in Eq. B42:

\begin{eqnarray}
\frac{\epsilon^3}{3} \left(  \sum_{i} \mu_i(1-\mu^2_i) \left< h^3_i \right> + 3 \sum_{i\ne j} \left(\pi_{ij} - \mu^2_i \mu_j \right) \left< h^2_i h_j  \right> + \sum_{i \ne j \ne k} \left( \tau_{ijk} - \mu_i \mu_j \mu_k \right) \left<h_i h_j h_k \right>   \right). 
\end{eqnarray}

What remains is to compute the $\epsilon^3$ term (with no factor of $1/3$), and we'll again do that term-by-term, noting that the first part ($\E_{\vec{h^s}}\left[   (\E_{SAD}[s])^3 \right]$) is already known (Eq. B51), leaving us to compute

\begin{eqnarray}
 \E_{\vec{h^s}} \left[  \E_{SAD}[s^2]  \E_{SAD} \left[ s \right] \right]   &=&  \E_{\vec{h^s}} \left[ \left( \sum_i \mu_i h^2_i + \sum_{i\ne j} \pi_{ij} h_i h_j \right) \sum_k \mu_k h_k \right]  \\
 &=&\E_{\vec{h^s}} \left[ \sum_i \mu^2_i h^3_i +  \sum_{i\ne j} \mu_i \mu_j h^2_i h_j + 2 \sum_{i\ne j} \mu_i \pi_{ij} h^2_i h_j +  \sum_{i\ne j \ne k} \mu_k \pi_{ij} h_i h_j h_k         \right]\nonumber \\
 &=&  \sum_i \mu^2_i  \left< h^3_i \right>+  \sum_{i\ne j} \mu_i \mu_j \left<h^2_i h_j \right> + 2 \sum_{i\ne j} \mu_i \pi_{ij} \left<h^2_i h_j \right> +  \sum_{i\ne j \ne k} \mu_k \pi_{ij} \left<h_i h_j h_k \right>      \nonumber
 \end{eqnarray}

where  we have used Eq. B44 for $ \E_{SAD}[s^2] $, and the factor of $2$ on the second line comes in because $k$ could be either $j$ or $i$. We can assemble this with Eq. B51 to get the $\epsilon^3$ term (with no factor of $1/3$) in our MI formula;

\begin{eqnarray}
&\epsilon^3 \left(   \sum_{i} \mu^2_i (\mu_i -1)  \left< h^3_i \right> + \sum_{i\ne j} \mu_i \left( 3\mu_i \mu_j  - \mu_j - 2 \pi_{ij} \right) \left<h^2_i h_j \right> + \sum_{i \ne j \ne k} \mu_k \left( \mu_i \mu_j - \pi_{ij} \right) \left< h_i h_j h_k\right>      \right) \nonumber \\
&=\epsilon^3 \left( -  \sum_{i} \mu_i \text{var}(\sigma_i) \left< h^3_i \right> + \sum_{i\ne j} \mu_i \left( \mu_i \mu_j  - \mu_j - 2 \text{cov}(\sigma_i, \sigma_j) \right) \left<h^2_i h_j \right> - \sum_{i \ne j \ne k} \mu_k \text{cov}(\sigma_i,\sigma_j) \left< h_i h_j h_k\right>      \right).   \nonumber
\end{eqnarray}

Assembling the pieces (above and Eqs. B42,B48,B52), we observe that

\begin{eqnarray}
MI &=&   \frac{\epsilon^2}{2} \left[     \sum_i  \nu_{ii} \text{var}(\sigma_i) + \sum_{i\ne j}  \nu_{ij} \text{cov}(\sigma_i,\sigma_i) \right]   \nonumber \\
&+& \frac{\epsilon^3}{3} \left(  \sum_{i} \mu_i(1-\mu^2_i) \left< h^3_i \right> + 3 \sum_{i\ne j} \left(\pi_{ij} - \mu^2_i \mu_j \right) \left< h^2_i h_j  \right> + \sum_{i \ne j \ne k} \left( \tau_{ijk} - \mu_i \mu_j \mu_k \right) \left<h_i h_j h_k \right>   \right) \nonumber \\
&+& \epsilon^3 \left( -  \sum_{i} \mu_i \text{var}(\sigma_i) \left< h^3_i \right> + \sum_{i\ne j} \mu_i \left( \mu_i \mu_j  - \mu_j - 2 \text{cov}(\sigma_i, \sigma_j) \right) \left<h^2_i h_j \right> - \sum_{i \ne j \ne k} \mu_k \text{cov}(\sigma_i,\sigma_j) \left< h_i h_j h_k\right>      \right).   \nonumber \\
&+& \mathcal{O}(\epsilon^4). \nonumber
\end{eqnarray}

We can simplify this a bit further by grouping together the $\epsilon^3$ terms, to yield

\begin{eqnarray}
MI &=&   \frac{\epsilon^2}{2} \left[     \sum_i  \nu_{ii} \text{var}(\sigma_i) + \sum_{i\ne j}  \nu_{ij} \text{cov}(\sigma_i,\sigma_i) \right]   \\
&+& \epsilon^3 \left(  \sum_{i} \frac{\mu_i}{3} \left(1 + 2 \mu^2_i - 3\mu_i \right) \left< h^3_i \right> +  \sum_{i\ne j} \text{cov}(\sigma_i,\sigma_j)\left(1 - 2 \mu_i \right) \left< h^2_i h_j  \right>  \right) \nonumber \\
&+& \epsilon^3 \left( \sum_{i \ne j \ne k} \frac{1}{3} \left( \tau_{ijk} + 2 \mu_i \mu_j \mu_k - 3 \mu_k \pi_{ij} \right) \left<h_i h_j h_k \right>   \right)  \nonumber \\
&+& \mathcal{O}(\epsilon^4). \nonumber
\end{eqnarray}

Recall that $\mu_{i} = \E_{SAD}(\sigma_i )$, $\pi_{ij} = \E_{SAD}(\sigma_i \sigma_j)$, $\tau_{ijk} = \E_{SAD}(\sigma_i \sigma_j \sigma_k)$, $\nu_{ij} = \E_{\vec{h^s}}(h_i h_j)$, angled brackets denote averages over the stimulus ensemble, and $\text{cov}(\sigma_i,\sigma_i) $ and $\text{var}(\sigma_i)$ are moments of the spontaneous activity distribution. Cleaning up our notation to match the main paper, we thus find that

\begin{eqnarray}
MI &=&   \frac{\epsilon^2}{2} \left[     \sum_i  \left< \left(h^s_i\right)^2 \right> \text{var}(\sigma_i) + \sum_{i\ne j} \left<h^s_i h^s_j \right> \text{cov}(\sigma_i,\sigma_i) \right]   \\
&+& \epsilon^3 \left[ \sum_{i} \frac{\mu_i}{3} \left(1 + 2 \mu^2_i - 3\mu_i \right) \left< h^3_i \right> +  \sum_{i\ne j} \text{cov}(\sigma_i,\sigma_j)\left(1 - 2 \mu_i \right) \left< h^2_i h_j  \right>  \right] \nonumber \\
&+& \epsilon^3 \left[ \sum_{i \ne j \ne k} \frac{1}{3} \left( \tau_{ijk} + 2 \mu_i \mu_j \mu_k - 3 \mu_k \pi_{ij} \right) \left<h_i h_j h_k \right>   \right]  \nonumber \\
&+& \mathcal{O}(\epsilon^4). \nonumber
\end{eqnarray}

\section{Numerical Methods}
\label{sec:num_meth}

\subsection{Monte Carlo methods and optimization}
The mutual information between the stimuli and responses,
\begin{eqnarray}
MI = - \sum_{ \{ \vec{\sigma} \}} p(\vec{\sigma}) \log[  p(\vec{\sigma})  ]  + \int d \vec{h^s} p(\vec{h^s}) \sum_{ \{ \vec{\sigma} \}} p(\vec{\sigma}| \vec{h^s}) \log[  p(\vec{\sigma}| \vec{h^s})  ],   
\end{eqnarray}
involves the sum over all $2^N$ possible population states, and an integral over the stimulus distribution of another such sum. This function is not analytically tractable for $N=10$ (the network size considered in this work) and / or for continuous stimulus distributions. Instead, we use Monte Carlo methods to compute the MI. In particular, we define the (un-normalized) frequency function
\begin{eqnarray}
\phi(\vec{\sigma} | \vec{h^s}) = \exp \left[\beta(  \vec{h^s}\cdot \vec{\sigma} +  {h^0}\sum_i {\sigma_i} +  J \sum_{i<j}  \sigma_i \sigma_j + \gamma \sum_{i<j<k} \sigma_i \sigma_j   \sigma_k   ) \right].  
\end{eqnarray}
This (log-polynomial) function can be very quickly evaluated, and to compute the MI, we take a large number of stimuli $\vec{h^s}$ from the appropriate distribution, and evaluate the frequencies of each of the $2^N$ states for each of the stimuli. We then divide the frequencies for each  state and stimulus by the sum of the frequencies over all states for that stimulus, to get (normalized) conditional probabilities:
\begin{eqnarray}
p(\vec{\sigma} | \vec{h^s}) =   \phi(\vec{\sigma} | \vec{h^s})  / \sum_{\{ \vec{\sigma} \}}   \phi(\vec{\sigma} | \vec{h^s}).       
\end{eqnarray}
This normalizing operation can be done quickly using matrix operations in MatLab~\cite{matlab}. Note that, if one instead defined the conditional probability for each state (instead of frequencies), then one would need to evaluate the partition function (costly) in the calculation of the probability of each of the $2^N$ states. Using the approach of first computing frequencies, we evaluate the partition function only once for each stimulus value, saving $2^N-1$ evaluations of the partition function for each stimulus example. Given the conditional probabilities, we then compute the conditional entropy (for each stimulus),
\begin{eqnarray}
H(\vec{\sigma} | \vec{h^s}) =   -\sum_{\{ \vec{\sigma} \}}   p(\vec{\sigma} | \vec{h^s}) \log[p(\vec{\sigma} | \vec{h^s})].  
\end{eqnarray}
Averaging these values over the set of stimuli from our distribution we get the noise entropy ($H_{noise}$, which is minus the second term in Eq.~C1). Similarly, we can average the conditional probabilities across all stimulus examples to get the (marginal) response distribution
\begin{eqnarray}
p(\vec{\sigma}) = \left< p(\vec{\sigma} | \vec{h^s}) \right>_{\vec{h^s}}.
\end{eqnarray}
Finally, we compute the entropy of the response distribution
\begin{eqnarray}
H_{resp} = - \sum_{\{ \vec{\sigma} \}}  p(\vec{\sigma}) \log[ p(\vec{\sigma}) ]
\end{eqnarray}
and subtract the noise entropy to get the MI: $MI = H_{resp} - H_{noise}$.

Note that, since we are using Monte Carlo integration, each evaluation of the MI function involves a (potentially) different set of stimuli, and thus a potentially (slightly) different result, even for identical network parameters. This noise makes gradient-based optimization methods highly error-prone. We avoid this pitfall by using exactly the same set of stimuli in subsequent calls to the MI function during the optimization. This \emph{common random number} approach makes the MI a smooth function of our parameters, allowing us to use gradient-based optimization techniques; see~\cite{deng} for an overview of optimization methods for noisy functions. For the optimization itself, we use the open-source MinFunc package~\cite{minfunc} from Mark Schmidt. We found that MinFunc was much faster and more reliable than the minimizers in the MatLab optimization toolbox. 

In this paper, we have used ensembles of $1000$ stimulus examples in evaluating the MI function. We repeated the optimization 5 times, with different sets of stimuli each time, and found that the results were highly reproducible: the standard deviation of the mean MI achieved over those 5 trials is small (Fig. 3A,B of the main paper) -- it is comparable to, or in many cases less than, the line width on the plots -- as is the standard deviation of the mean parameter values obtained at optimality (Figs. 4A,B of the main paper).

The expressions herein (and in the main paper) do not specify the base in which the logarithm is computed. For MI values in bits, those logarithms are to base 2. 

\subsection{Comparing optimal networks with constrained firing rates}
When we use Lagrange multipliers for optimizing MI with constrained firing rates (see main paper), the exact functional relationship between Lagrange multiplier $\lambda$ and firing rate is unknown: although higher Lagrange multipliers lead to lower firing rates, we cannot easily specify what value of $\lambda$ is needed to achieve a given firing rate. We use the same values of the Lagrange multipliers when we optimize with triplet interactions either allowed (TA), or forbidden (TF), resulting in (slightly) different mean firing rates for the optimal TA and TF networks. The reason for this difference is easy to understand, as they have different MI values, and thus the optimal trade-off between MI and firing rate in the Lagrange function $\mathcal{L} = MI - \lambda \left< \sigma \right>$ will be slightly different.

We use linear interpolation to estimate the MI of the TF network at the \emph{exact mean firing rate of the TA network}: since we have several points on the curve of MI vs. mean firing rate for the TF network, this interpolation is easy to implement.
Finally, we take the ratio of the MI value for the TA network to the (interpolated) one for the TF network at the same firing rate to create the data in Figs. 3E,F.

\section{Triplet interactions are better than pair-wise ones at suppressing multi-spike states, hence the observation that $\gamma< 0$ and $J > 0$ at optimality, and not vice versa}

Consider the contribution $C$ of the recurrent connections to the log-polynomial probability distribution over network states, $C = J \sum_{i<j}  \sigma_i \sigma_j + \gamma \sum_{i<j<k} \sigma_i \sigma_j   \sigma_k$. When this number is large, the state is favored, and vice versa. The first term ($J \sum_{i<j}  \sigma_i \sigma_j $) is $J$ times the number of active neural pairs, which is $\mathcal{O}(\alpha^2)$, where $\alpha$ is the number of co-active neurons, while the second term ($\gamma \sum_{i<j<k} \sigma_i \sigma_j   \sigma_k$) is $\mathcal{O}(\alpha^3)$.

Let us consider situations in which it is desirable for neurons act cooperatively, while not always firing synchronously.

If we choose positive $J$ and (small) negative $\gamma$ -- which is what we observe at optimality: see Fig. 4B of the main paper -- then neurons are encouraged to be co-active by the positive $J$: they cooperate. If one considers states with many spikes (large $\alpha$), however, we can see that the effects of the triplet interaction, which are $\mathcal{O}(\alpha^3)$, can exceed the pair-wise ones, which are $\mathcal{O}(\alpha^2)$. In other words, $C \sim J \mathcal{O}(\alpha^2) + \gamma \mathcal{O}(\alpha^3)$ is a unimodal function with a positive peak, such that for large $\alpha$, states are strongly suppressed, while for intermediate $\alpha$, they may be facilitated (Fig. 5: upper (pink) curve). This acts somewhat like negative feedback: for small $\alpha$, the effects of $J$ (larger in magnitude than $\gamma$) dominate, pushing the network towards having co-active cells, while for large $\alpha$, the effects of $\gamma$ push the network away from having too many co-active cells. Thus, the network produces cooperative responses but has a diminished probability of having all of the neurons co-active.

Now consider the opposite situation, with negative $J$, and positive $\gamma$. In this case $C$ is unimodal with a negative peak, and the larger-$\alpha$ states are progressively more facilitated by recurrent interactions (Fig. 5: lower (brown) curve). This is reminiscent of positive feedback, and leads to heavy usage of the all-neurons-on state. 

Of course, if we allow $4^{th}$ order terms in the probability model, then one could have positive $\gamma$, while still avoiding epileptic levels of synchrony, by having negative  $4^{th}$ order interactions, for example.

%%%%%%%%%Figure%%%%%%%%%%%%%%%%%%
\begin{figure}[tb!]
\begin{center}
\includegraphics[width=4.0in]{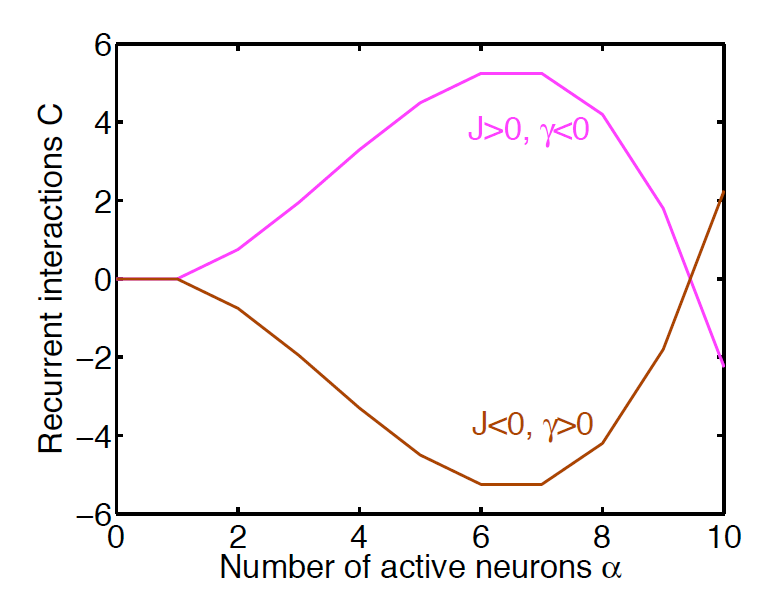}
\caption{(Color online) \textbf{Negative triplet interactions are better at suppressing the all-on state than are negative pairwise interactions.} The plots here show the function $C = J \alpha (\alpha-1)/2 + \gamma  \alpha (\alpha-1) (\alpha-2)/6$ for non-negative integer values of $\alpha \in [0,10]$. For J=0.75 and $\gamma=-0.3$ (upper, pink curve), the states with many co-active neurons are suppressed by the recurrent interaction. For J=-0.75 and $\gamma=0.3$ (lower, brown curve), the opposite is true.}
\end{center}
\end{figure}
%%%%%%%%%Figure%%%%%%%%%%%%%%%%%%


\begin{thebibliography}{99}

\bibitem{stevenson}
I.H. Stevenson and K.P. Kording, Nat. Neurosci. \textbf{14}, 139 (2011).

\bibitem{schneidman06}
E. Schneidman, M.J. Berry, R. Segev, and W. Bialek, Nature  \textbf{440}, 1007 (2006).

\bibitem{shlens0609}
J. Shlens et al., J. Neurosci. 26, 8254 (2006).; ibid. J. Neurosci.  \textbf{15}, 5022 (2009).

\bibitem{granot13}
E. Granot-Atedgi, G. Tka\v{c}ik, R. Segev, and E. Schneidman, PLoS Comput. Biol \textbf{9}, e1002922 (2013).

\bibitem{ganmor}
E. Ganmor, R. Segev, and E. Schneidman, Proc. Natl. Acad. Sci. USA  \textbf{108}, 9679 (2011).

\bibitem{tkacik14}
G. Tkacik, O. Marre, , D. Amodei, E. Schneidman, W. Bialek, and M.J. Berry II, PLoS Comp. Biol.  \textbf{10}, e1003408 (2014).

\bibitem{montani}
F. Montani \emph{ et al.}, Phil. Trans. R. Soc. A   \textbf{367}, 3297-3310 (2009).

\bibitem{yu}
S. Yu, H. Yang, H. Nakahara, G. Santos, D. Nikoli\'{c}, and D. Plenz, J. Neurosci.  \textbf{31}, 17514 (2011).

\bibitem{ohiorhenuan}
I.E. Ohiorhenuan, F. Mechler, K.P. Purpura, A.M. Schmid, and J.D. Victor, Nature  \textbf{466}, 617  (2010).

\bibitem{averbeck}
B. Averbeck, P.E. Latham, and A. Pouget, Nat. Rev. Neurosci.  \textbf{7}, 358 (2006).

\bibitem{cafaro}
J. Cafaro and F. Rieke, Nature \textbf{468}, 964 (2010).

\bibitem{roudi_latham}
P.E. Latham and Y. Roudi, arXiv:1109-6524v1 (2011).

\bibitem{zohary}
E. Zohary, M.N. Shadlen, and W.T. Newsome, Nature \textbf{370}, 140 (1994).

\bibitem{tkacik}
G. Tka\v{c}ik, J.S. Prentice, V. Balasubramanian, and E. Schneidman, Proc. Natl. Acad. Sci. USA \textbf{107}, 14419 (2010).

\bibitem{beck}
J. Beck, V.R. Bejjanki, and A. Pouget, Neural Comput. \textbf{23}, 1484-1502 (2011).

\bibitem{ecker11}
A.S. Ecker, P. Behrens, A.S. Tolias, and M. Bethge, J. Neurosci. \textbf{31}, 14272-14283  (2011).

\bibitem{sha04}
M. Shamor and H. Sompolinsky, Neural Comput. \textbf{16}, 1105 (2004).

\bibitem{josic09}
K. Josic, E. Shea-Brown, J. de la Rocha, and B. Doiron, Neural Comput. \textbf{21}, 2774 (2009). 

\bibitem{hu14}
Y. Hu, J. Zylberberg, and E. Shea-Brown, PLoS Comput. Biol. \textbf{10}, e1003469 (2014).

\bibitem{cayco15}
N.A. Cayco-Gajic, J. Zylberberg, and E. Shea-Brown, Frontiers Comput. Neurosci. \textbf{9}, 57 (2015).


\bibitem{salakhutdinov09}
R. Salakhutdinov and G.E. Hinton, ``Deep Boltzmann Machines". Proc. Intl. Conf. Artificial Intelligence and Statistics \textbf{5}, 448 (2009).

\bibitem{hinton83}
G.E. Hinton, and T.J. Sejnowski, ``Optimal Perceptual Inference". Proc. IEEE Comput. Vis. Pattern Recog, 448 (1983). 

\bibitem{betacomment}
 Our $\beta$ parameter is effectively equivalent to double that of Tka\v{c}ik et al.~\cite{tkacik}: we use neurons with $\sigma_i \in \{0,1\}$, and they use ones with $\sigma_i \in \{-1,1\}$. The conversion is $\sigma^{\{-1,1\}} = 2(\sigma^{\{0,1\}} -1/2)$.

\bibitem{schneidman_connected}
E. Schneidman, S. Still, M.J. Berry, and W. Bialek, Phys. Rev. Lett.  \textbf{91}, 238701 (2003).

\bibitem{jaynes57}
E.T. Jaynes, Phys. Rev. \textbf{106}, 620 (1957).

\bibitem{amari01}
S.-I. Amari, IEEE Trans. Info. Theory \textbf{47}, 1701 (2001).

\bibitem{maass99}
W. Maass, Computing with spiking neurons. In: W. Maass and C. M. Bishop, eds. Pulsed Neural Networks. Cambridge: MIT Press.  pp. 55-85  (1999). 

\bibitem{brillinger}
D. R. Brillinger and J.P. Segundo, Biol. Cybern. \textbf{35}, 213 (1979).

\bibitem{ho2000}
N. H\^{o} and A. Destexhe, J. Neurophysiol. \textbf{84}, 1488 (2000).


\bibitem{macke}
J.H. Macke, M. Opper, and M. Bethge, Phys. Rev. Lett. \textbf{106}, 208102 (2011).

\bibitem{andrea}
A.K. Barreiro, J. Gjorjieva, F. Rieke, and E. Shea-Brown, Frontiers Comput. Neurosci. \textbf{8}, 10 (2014).

\bibitem{expansion_comment}
We use this re-write so that we can expand the MI in powers of the stimulus-coupling strength. This expansion involves taking derivatives of the MI with respect to the stimulus-coupling parameter. In the case where the same coupling parameter multiplies the stimulus strength and the interaction parameters, this expansion becomes much more involved. 

\bibitem{leen}
 D. Leen and E. Shea-Brown, J. Math. Neurosci. \textbf{5}, 17 (2015).

\bibitem{barlow}
H. B. Barlow, In W. A. Rosenblith, editor, Sensory Communication, pp. 217-234. Cambridge: MIT Press (1961).

\bibitem{simoncelli_olshausen}
E.P. Simoncelli and B.A. Olshausen, Annu. Rev. Neurosci. \textbf{24}, 1193 (2001).

\bibitem{skew_comment}
We note that, if the photoreceptors perform a logarithmic transformation on the luminance values that they sample from the environment, then the downstream inputs (say, to retinal ganglion cells) may have much less skewed distributions than do the raw luminance values. In that case, our arguments about skewed inputs may not apply directly to retinal ganglion cells. At the same time, if there are firing rate constraints on retinal ganglion cells, we expect that beyond-pairwise correlations could enhance their information coding ability regardless of the skewness of their inputs (Fig. 3E).

\bibitem{tkacik_images}
G. Tka\v{c}ik et al., PLoS One \textbf{6}, e20409 (2011).

\bibitem{ruderman_scaling}
D.L. Ruderman and W. Bialek, Phys. Rev. Lett. \textbf{73}, 814  (1994).

\bibitem{stephens_image_statmech}
G.J. Stephens, T. Mora, G. Tkacik, and W. Bialek, Phys. Rev. Lett. \textbf{110}, 018701 (2013). 

\bibitem{ruderman97}
D. Ruderman, Vision Res. \textbf{37}, 3385 (1997).

\bibitem{hsiao}
Hsiao, W.H. and  Milane, R.P., J. Opt. Soc. Am. A \textbf{22}, 1789 (2005). 

\bibitem{jz_pfau}
J. Zylberberg, D. Pfau, and M.R. DeWeese, Phys. Rev. E \textbf{86}, 066112 (2012) [arXiv:1209.3277].

\bibitem{deweese_zador}
M.R. DeWeese and A.M. Zador (2006). J. Neurosci. 26: 12206-12218.

\bibitem{hromadka}
T. Hrom\'{a}dka, M.R. DeWeese, and A.M. Zador, PLoS Biol. \textbf{6}, e16 (2008).

\bibitem{baddeley}
R. Baddeley et al., Proc. R Soc. Lon. B \textbf{264}, 1775 (1997).

\bibitem{karklin_simoncelli}
Y. Karklin and E.P. Simoncelli, In J. Shawe-Taylor et al., eds, Advances in Neural Information Processing Systems, volume 24. Cambridge: MIT Press (2011). 

\bibitem{jz_plos}
J. Zylberberg, J.T. Murphy, and M.R. DeWeese, PLoS Comput. Biol. \textbf{7}, e1002250 (2011).

\bibitem{olshausen96}
B.A. Olshausen and D.J. Field, Nature \textbf{381}, 607 (1996).

\bibitem{kzd}
P.D. King, J. Zylberberg, and M.R. DeWeese, J. Neurosci. \textbf{33}, 5475 (2013).

\bibitem{jz_sailsparse}
J. Zylberberg and M.R. DeWeese, PLoS Comput. Biol. \textbf{8}, e1003182 (2013).

\bibitem{dayan_abbott}
P. Dayan and L.F. Abbott, Theoretical Neuroscience. Cambridge: MIT Press (2001). 

\bibitem{MEcomment}
We note that, for a given set of moments, the pairwise maximum entropy model (with $\gamma = 0$) maximizes the entropy of the responses. Thus, it may seem puzzling to think of the inclusion of $\gamma \ne 0$ increasing response entropy. In this situation here, we are not holding the response statistics fixed. The optimization procedure co-varies $J$ and $\gamma$ to maximize the MI, but the response statistics (correlations, etc.) can vary. Consequently, it is possible for optimal $\gamma \ne 0$ models (triplet allowed) to have larger response entropy than optimal $\gamma = 0$ (triplet forbidden) ones.

\bibitem{gulledge}
A.T. Gulledge, B.M. Kampa, and G.J. Stuart, J. Neurobiol. \textbf{64}, 75 (2005).

\bibitem{mel}
B.W. Mel, J. Neurophysiol. \textbf{70}, 1086 (1993).

\bibitem{harnett}
M.T. Harnett, J.K. Makara, N. Spruston, W.L. Kath, and J.C. Magee, Nature \textbf{491}, 599  (2012).

\bibitem{london}
M. London and M. H\"{a}usser, Annu. Rev. Neurosci. \textbf{28}, 503 (2005).

\bibitem{polsky}
A. Polsky, B.W. Mel, and J. Schiller, Nat. Neurosci. \textbf{7}, 622 (2004).

\bibitem{urs}
U. Koster, J. Sohl-Dickstein, C.M. Gray, and B.A. Olshausen, PLoS Comput. Biol. \textbf{10} e1003684 (2014).

\bibitem{amari03}
S.-I. Amari, H. Nakahara, S. Wu, and Y. Sakai, Neural Comput. \textbf{15}, 127 (2003).

\bibitem{matlab}
 MATLAB version 2012a Natick, Massachusetts: The MathWorks Inc. (2012).
 
 \bibitem{deng}
G. Deng, Simulation-based optimization techniques. PhD Thesis, University of Wisconsin, Madison (2007).

\bibitem{minfunc}
M. Schmidt, minFunc: unconstrained differentiable multivariate optimization in Matlab. \underline{http://www.cs.ubc.ca/~schmidtm/Software/minFunc.html} (2005).








%\bibitem{olshausen96}
%Olshausen, B.A. and Field, D.J. (1996). Nature. 381, 607-609.

%\bibitem{rehn_sommer}
%Rehn, M. and Sommer, F.T. (2007). J Comput. Neurosci. 22, 135-146.



\end{thebibliography}
\end{document}